\documentclass[twocolumn,showpacs,preprintnumbers,amsmath,amssymb,superscriptaddress]{revtex4-1}

\usepackage{graphicx,footnote}
\usepackage{epstopdf} 
\usepackage{enumerate}
\usepackage[format=plain,singlelinecheck=off]{caption}
\usepackage{subcaption}
\usepackage{ngerman}
\usepackage[utf8]{inputenc}
\usepackage{amsmath}
\usepackage[stable]{footmisc}
\usepackage{dsfont}
\usepackage{textcomp}
\usepackage{listliketab}
\usepackage{tabularx}
\usepackage{multirow}
\usepackage{color}
\usepackage{tabto}
\usepackage{stmaryrd}
\usepackage{amssymb}
\usepackage{here}
\usepackage{footnote}
\usepackage{pstricks,pstricks-add}
\usepackage{pst-plot}
\usepackage{tikz}
\usetikzlibrary{arrows,shapes}
\usepackage{extarrows}
\usepackage{hyperref}

\makeatletter 
\g@addto@macro\normalsize{%
\setlength{\abovedisplayskip}{5pt} 
\setlength{\belowdisplayskip}{10pt}} 
\makeatother

\newcommand*\teo{  \begin{minipage}[H]{5cm}
\begin{tikzpicture}[scale=1] 
\draw (0,0) -- (5,0); \draw [ultra thick] (0,-0.25) -- (0,0.25); \draw (1.6,-0.25) -- (1.6,0.25); \draw (1.66,-0.25) -- (1.66,0.25); \draw (2.6,-0.25) -- (2.6,0.25); \draw (2.66,-0.25) -- (2.66,0.25); \draw [fill] (4.9,0.08) -- (4.9,-0.08) -- (5,0) -- (4.9,0.08);  \draw [fill]  (1.2,0)circle (2pt); \draw [ultra thick] (1.2,0) -- (4.9,0);
\end{tikzpicture}
\end{minipage}}

\newcommand*\mbo{ \begin{minipage}[H]{5cm}
\begin{tikzpicture}[scale=1] 
\draw (0,0) -- (5,0); \draw [ultra thick] (0,-0.25) -- (0,0.25); \draw (1.6,-0.25) -- (1.6,0.25); \draw (1.66,-0.25) -- (1.66,0.25); \draw (2.6,-0.25) -- (2.6,0.25); \draw (2.66,-0.25) -- (2.66,0.25); \draw [fill] (4.9,0.08) -- (4.9,-0.08) -- (5,0) -- (4.9,0.08); \draw [fill]  (1.2,0)circle (2pt); \draw [fill]  (3.06,0)circle (2pt); \draw [ultra thick] (1.2,0) -- (3.06,0); 
\end{tikzpicture}
\end{minipage}}
			
\newcommand*\tos{ \begin{minipage}[H]{5cm}
\begin{tikzpicture}[scale=1]
\draw (0,0) -- (5,0); \draw [ultra thick] (0,-0.25) -- (0,0.25); \draw (1.6,-0.25) -- (1.6,0.25); \draw (1.66,-0.25) -- (1.66,0.25); \draw (2.6,-0.25) -- (2.6,0.25); \draw (2.66,-0.25) -- (2.66,0.25);\draw [fill] (4.9,0.08) -- (4.9,-0.08) -- (5,0) -- (4.9,0.08); \draw[fill] (0,-2pt) arc (-90:90:2pt);  \draw [ultra thick] (0,0) -- (4.9,0); 
\end{tikzpicture}
\end{minipage}}
					
\newcommand*\mboeo{ \begin{minipage}[H]{5cm}
\begin{tikzpicture}[scale=1]
\draw (0,0) -- (5,0); \draw [ultra thick] (0,-0.25) -- (0,0.25); \draw (1.6,-0.25) -- (1.6,0.25); \draw (1.66,-0.25) -- (1.66,0.25); \draw (2.6,-0.25) -- (2.6,0.25); \draw (2.66,-0.25) -- (2.66,0.25); \draw [fill] (4.9,0.08) -- (4.9,-0.08) -- (5,0) -- (4.9,0.08); \draw [fill]  (1.2,0)circle (2pt); \draw [fill]  (3.8,0)circle (2pt); \draw [fill]  (3.06,0)circle (2pt); \draw [ultra thick] (1.2,0) -- (3.06,0);  \draw [ultra thick] (3.8,0) -- (4.9,0);
\end{tikzpicture}
\end{minipage}}
							
\newcommand*\boteo{ \begin{minipage}[H]{5cm}
\begin{tikzpicture}[scale=1]
\draw (0,0) -- (5,0); \draw [ultra thick] (0,-0.25) -- (0,0.25); \draw (1.6,-0.25) -- (1.6,0.25); \draw (1.66,-0.25) -- (1.66,0.25); \draw (2.6,-0.25) -- (2.6,0.25); \draw (2.66,-0.25) -- (2.66,0.25); \draw [fill] (4.9,0.08) -- (4.9,-0.08) -- (5,0) -- (4.9,0.08); \draw [fill]  (0.3,0)circle (2pt); \draw [fill]  (0.8,0)circle (2pt); \draw [fill]  (1.2,0)circle (2pt); \draw [ultra thick] (0.3,0) -- (0.8,0);  \draw [ultra thick] (1.2,0) -- (4.9,0);
\end{tikzpicture}
\end{minipage}}

\begin{document}

\title{Dynamics of test particles in the five-dimensional, charged, rotating EMCS spacetime}

\author{Shruti Paranjape}
\email{shrpar@umich.edu}
\affiliation{Indian Institute of Science Education and Research, Pune, Maharashtra 411021, India}

\author{Stephan Reimers}
\email{stephan.reimers@uni-oldenburg.de}
\affiliation{Institut f\"{u}r Physik, Universit\"{a}t Oldenburg, 26111 Oldenburg, Germany}

\date{\today}

\begin{abstract}
\noindent
We derive the complete set of geodesic equations for massive and massless test particles of a five-dimensional, charged, rotating black hole solution of the Einstein-Maxwell-Chern-Simons field equations in five-dimensional minimal gauged supergravity and present their analytical solutions in terms of Weierstraß’ elliptic functions. We study the polar and radial motion, depending on the black hole and test particle parameters, and characterize the test particle motion qualitatively by the means of effective potentials. We use the analytical solutions in order to visualize the test particle motion by two- and three-dimensional plots.
\end{abstract}

\maketitle

\section{Introduction}

All regular, stationary, asymptotically flat solutions of the Einstein-Maxwell field equations in $D=4$ dimensions are uniquely determined by their mass, angular momentum, and electric charge, and are in fact given by the Kerr-Newman family of solutions. Furthermore, this family may be extended by a magnetic charge, a cosmological constant, a Taub-NUT charge \cite{Newman:1963yy} and an accalaration parameter yielding the complete family of Petrov type D spacetimes. In the asymptotically flat vacuum case, this family reduces to the Kerr family, being determined only by a mass and an angular momentum parameter \cite{Kerr:1963ud}. In 1986, R. Myers and M. Perry generalized the Kerr solution to higher dimensions \cite{Myers:1986un}. Depending on the number of dimensions $D$, the Myers-Perry solutions possess $N=(D-1)/2$ independent
angular momenta, associated with rotation in $N$ orthogonal planes.\\
Remarkably, five-dimensional, stationary vacuum black holes are not unique. Besides the Myers-Perry solution, Emparan and Reall have found five-dimensional rotating black ring solutions \cite{Emparan:2001wn} with the same angular momenta and mass but now a non-spherical event horizon topology.\\ 
Further generalizations of the Myers-Perry solutions include the general Kerr-de Sitter and Kerr-NUT-AdS metrics in all higher dimensions \cite{Gibbons:2004uw,Chen:2006xh}.\\
Today, string theory is a promising candidate for the quantum theory of gravity. Since it requires extra dimensions of spacetime for their mathematical consistency, there has been growing interest in higher-dimensional solutions and, in particular,
in higher-dimensional black hole solutions.\\ The geodesic equation is a powerful tool used to discuss the properties of a spacetime and exact solutions of the geodesic
equations can be used to calculate spacetime observales to arbitrary accuracy. There is further interest to understand explicitly the structure of
geodesics in the background of black holes in anti-de Sitter space in the context of string theory and the AdS/CFT correspondence \cite{Maldacena:1997re}.\\
However, the four-dimensional Kerr-Newman solution of Einstein's field equation couldn't be generalized to higher dimensions, yet. Nevertheless, there is a related solution of the Einstein-Maxwell-Chern-Simons (EMCS) equations of motion in the five-dimensional minimal gauged supergravity \cite{Chong:2005hr, Gauntlett:2002nw}. This solution is determined by the mass, two angular momenta, an eletric charge and the cosmological constant. A special case of this general spacetime has been found earlier by J. Breckendridge, R. Myers A. Peet and C. Vafa \cite{Breckenridge:1996is}. This so-called BMPV spacetime describes the extremal case with equal-magnitude angular momenta \cite{Gauntlett:1998fz} . The analytical solutions of the related geodesic equations have been investigated in \cite{Gibbons:1999uv, Herdeiro:2000ap} as well as in \cite{Diemer:2013fza}.\\
Reviews of higher-dimensional black hole solutions in vacuum or in supergravity theory are, for instance, found in \cite{Emparan:2008eg,Emparan:2008zz}. Here also black objects with non-spherical horizon topology are discussed together with the associated non-uniqueness of higher-dimensional black holes.\\
In this paper, we want to explore the geodesic features of the rotating, asymptotically flat solution of the Einstein-Maxwell-Chern-Simons equations in five-dimensional minimal gauged supergravity.\\
In Sec. II, we will present the basic features of this spacetime and derive the geodesic equation by solving the Hamilton-Jacobi equation. Sec. III contains a qualitative discussion and a complete characterization of the test particle dynamics, especially the radial effective potentials are introduced. Sec. IV is dedicated to the analytical solutions of the equations of motions obtained in Sec. II, which will be used in Sec V in order to calculate integral expressions for selected spacetime observales. Finally, we will present two- and three-dimensional representations of the related orbits in Sec VI.

\section{The five-dimensional, charged, rotating EMCS spacetime}

We will briefly recall the basic properties of the five-dimensional, rotating, asymptotically flat solution of the Einstein-Maxwell-Chern-Simons equations and derive the geodesic equations describing the  motion of massive and massless test particles.

\subsection{Metric}

The bosonic sector of minimal $D=5$ supergravity
is described by the Lagrangian \cite{Cremmer:1981}

\begin{align}
{\cal L}= \frac{1}{16\pi } \left[\sqrt{-g}(R -F^2) -
\frac{2}{3\sqrt{3}}\epsilon^{\mu \nu \rho \sigma \tau}A_\mu F_{\nu \rho}F_{\sigma \tau}\right],
\end{align}

with curvature scalar $R$, 
gauge potential $A_\mu$,
field strength tensor
$ F_{\mu \nu} = \partial_\mu A_\nu -\partial_\nu A_\mu $,
and the five-dimensional Levi-Civita tensor density 
$\epsilon^{\mu \nu \lambda \rho \sigma}$ with $\epsilon^{01234} = -1$.
Thus, besides the usual Maxwell term,
it includes the `$AFF$' Chern-Simons term with a particular coefficient
\cite{Cremmer:1981}.
Since this term is odd in the gauge field, it breaks
the $A \to -A$ invariance of pure Maxwell theory.

The metric and the one-form gauge field describing the 
five-dimensional, rotating, charged asymptotically flat black hole spacetime 
can be obtained from the Einstein-Maxwell-Chern-Simons (EMCS) 
equations of motion \cite{Gauntlett:2002nw}
\begin{align}
\begin{aligned}
R_{\mu \nu} - \frac{1}{2} g_{\mu \nu} R = 2 \left(F_{\mu \alpha} F_\nu{}^\alpha - \frac{1}{4} g_{\mu \nu} F_{\rho \sigma} F^{\rho \sigma} \right),\\
\nabla_\mu \left(F^{\mu \nu} + \frac{1}{\sqrt{3} \sqrt{-g}} \epsilon^{\mu \nu \lambda \rho \sigma} A_\lambda F_{\rho \sigma}\right) = 0.
\label{eq:EMCS}
\end{aligned}
\end{align}

The black hole solution is given by the metric

\begin{align}
\begin{aligned}
ds^2 = &- \frac{\rho^2 dt^2 + 2q\nu dt}{\rho^2} + \frac{2q \nu \omega}{\rho^2} + \frac{\mu \rho^2 - q^2}{\rho^4} \left(dt - \omega \right)^2\\ & + \frac{\rho^2 dx^2}{4\Delta} + \rho^2 d\theta^2 + \left(x+a^2 \right) \sin^2 \theta \, d\phi^2\\ &+ \left(x+b^2 \right) \cos^2 \theta \, d\psi^2
\label{eq:linelement}
\end{aligned}
\end{align}

with

\begin{align}
\begin{aligned}
\rho &= x + a^2 \cos^2 \theta + b^2 \sin^2 \theta,\\
\Delta &= \left(x+a^2 \right)\left(x+b^2 \right) + q^2 +2 abq - \mu x,\\
\nu &= b \sin^2 \theta \, d\phi + a \cos^2 \theta \, d\psi,\\
\omega &= a \sin^2 \theta \, d\phi + b \cos^2 \theta \, d\psi
\end{aligned}
\end{align}

in asymptotically static Boyer-Lindquist-like coordinates $(t,x,\theta,\phi,\psi)$ \cite{Boyer:1966qh} and the one-form gauge field

\begin{align}
A_\mu dx^\mu = \frac{\sqrt{3} \, q}{\rho^2} \left(dt - \omega \right), 
\label{eq:gaugefield}
\end{align}

where $m$ is the mass, $q$ is the charge and $a,b$ are the two independent angular momenta of the black hole.\\

The special case of $q=0$ yields the usual Myers-Perry solution of the Einstein field equations. Note that the standard radial Boyer-Lindquist coordinate $r$ has been subsituted in favor of a new radial coordinate $x$ via $x= r^2$, as proposed for any odd-dimensional Myers-Perry spacetime \cite{Myers:1986un}. The geodesic features of this spacetime have been studied in \cite{Kagramanova:2012hw} for equal-valued rotation parameters and in \cite{Diemer:2014lba} for the general case. The metric determinant is the same as for the uncharged case \cite{Myers:1986un}

\begin{align}
\sqrt{-\det g} = \frac{1}{2}\rho^2 \sin \theta\, \cos \theta.
\end{align}

By the form of the metric, we can see that it becomes singular at $\rho=0$ and $\Delta=0$, of which only $\rho=0$ is a physical singularity. The mere coordinate singularity $\Delta = 0$ defines the horizons of this spacetime, which are given by

\begin{align}
x_{\pm} = \frac{1}{2} \left(\mu - a^2 - b^2 \pm \sqrt{\left(\mu - a^2 - b^2 \right)^2 - 4\left(ab+q \right)^2} \right),
\end{align}

depending on the charge $q$. Here, $x^+$ defines the event horizon and $x^-$ is the Cauchy horizon. Both horizons merge in the extremal case of $\vert \mu - a^2 - b^2\vert = \vert 2 ab+q\vert$. 

\begin{figure}[htbp]
\centering
   \includegraphics[width=0.5\textwidth]{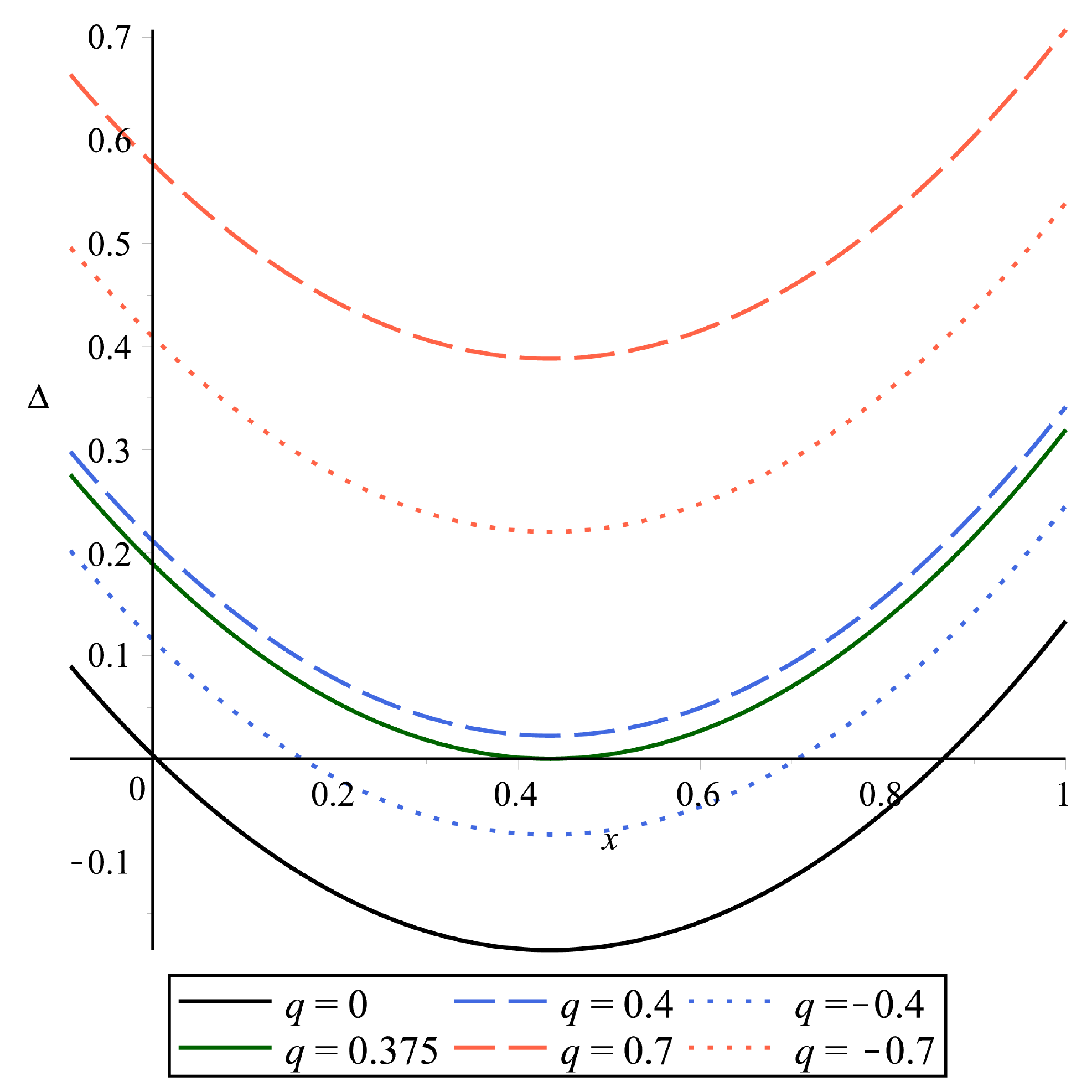}
\caption{Plot of $\Delta$ for $\mu=1$, $a=0.3$, $b=0.2$ and various values of $q$. The black solid line denotes the neutral case as well as dashed lines refer to positive charges $q$ and dotted lines refer to negative charges. The solid green line denotes the extremal case, where both horizons merge.}
\label{fig:horizons}
\end{figure}

The black hole's charge affects the constant term of $\Delta$, but is sign-dependent. So, while one sign yields two regular horizons, the other sign may lead to a naked singularity and must be forbidden by the cosmic censorship conjecture. This is different to the four-dimensional Kerr-Newman spacetime, where the horizons are the same for opposite signs of the black hole's charges. In the charged, rotating EMCS spacetime this degeneracy is removed as a result of the Chern-Simons term in the field equations.\\

The singularity at $\rho = 0$ will determine the curvature singularity, which is independent of the charge $q$. It is well-known that the Kerr and Kerr-Newmann curvature is ring-shaped. This can easily be seen in Kerr-Schild coordinates. Moreover, its circumference can by determined to be $2\pi a$, supporting this statement. In our case, we obtain the same singularity, which is present in the uncharged Myery-Perry spacetime. This curvature singularity is not ring-like but a non-traversable surface between the radial values $x=-b^2$ and $x=-a^2$, depending on the value of $\theta$ \cite{Diemer:2014lba}.\\

Another interesting feature of the charged, rotating EMCS spacetime is the static limit as well as the enclosed ergosphere, since it is an important hypersurface when considering frame-dragging effects. We define the static limit by demanding $g_{tt}(r,\theta)$ to be zero, so that the sign of $g_{tt}$ changes when this hypersurface is crossed. Analogous to the Kerr-Newman spacetime, this equation yields two solutions given by the relation

\begin{align}
x &=  \frac{1}{2} \left(\mu \pm \sqrt{\mu^2 - 4q^2}\right) - a^2 \cos^2\theta - b^2 \sin^2 \theta.
\end{align}

Obviously, the positive sign solution describes a hypersurface outside the event horizon, whole the negative sign solution is completely covered by the Cauchy horizon. While in the Kerr-Newman spacetime the corresponding horizons and static limits meet at the poles $(\theta=0,\theta = \frac{\pi}{2})$, this is only possible for certain sets of parameters in the charged, rotating EMCS spacetime. If we choose, e.g.,

\begin{align}
q = -\frac{ab\mu}{a^2 + b^2},
\end{align}

these points are given by

\begin{align}
\theta  = \arccos \left(\pm\sqrt{\frac{\mu}{a^2 + b^2}}\right).
\end{align}

Both solutions merge, when we choose $q=0$ for the uncharged case, leaving the single solution

\begin{align}
x &=  \mu - a^2 \cos^2\theta - b^2 \sin^2 \theta
\end{align}

for the Myers-Perry spacetime. The static limits of the four-dimensional analogon, given by the Kerr spacetime, do not merge for any non-zero rotation parameter. However, in the five-dimensional, charged, rotating EMCS spacetime, only a non-vanishing charge gives rise to both solutions of the static limit.

\subsection{Hamilton-Jacobi equation}

We will derive the geodesic equations of this spacetime by applying the Hamilton-Jacobi formalism. Therefore, we seek for a solution $S$ of the Hamilton-Jacobi equation \cite{Misner:1974qy}

\begin{align}
\label{eq:HamJac}
- \frac{\partial S}{\partial \lambda} = \frac{1}{2} g^{\mu \nu}\frac{\partial S}{\partial x^\mu}\frac{\partial S}{\partial x^\nu}.
\end{align}

Since the metric is independent of $t$, $\phi$ and $\psi$, we will relate the corresponding conjugate momenta to the test particle's energy $E$ and angular momenta $\Phi$ and $\Psi$, respectively. This motivates an ansatz of the form

\begin{align}
\label{eq:Sansatz}
S = \frac{1}{2} \delta \lambda - Et + S_r(r) + S_\theta(\theta) + \Phi \phi + \Psi \psi,
\end{align} 

where $S_r(r)$ and $S_\theta(\theta)$ are functions on $r$ and $\theta$ only. We introduced a mass parameter $\delta$ ($\delta = 1$ for massive and $\delta = 0$ for massless test particles) and an affine parameter $\lambda$.\\
The elements of the contravariant metric tensor are given by

\begin{align}
\begin{aligned}
g^{tt} &=\frac{1}{\rho^2} \left[(a^2-b^2)\sin^2 \theta - \frac{\alpha\beta \mu + \alpha \Delta - q^2(\alpha+b^2)}{\Delta}\right],\\
g^{xx} &=\frac{4 \Delta}{\rho^2},\\
g^{\theta \theta} &=\frac{1}{\rho^2},\\
g^{\phi \phi} &=\frac{1}{\rho^2} \left[\frac{1}{\sin^2\theta} - \frac{(a^2-b^2)\beta + b^2 \mu + 2abq}{\Delta} \right],\\
g^{\psi \psi} &=\frac{1}{\rho^2} \left[\frac{1}{\cos^2 \theta} - \frac{(b^2-a^2)\alpha + a^2 \mu + 2abq}{\Delta} \right],\\
g^{t \phi} &=\frac{(\beta\mu - q^2)a + b \beta q}{\rho^2 \Delta},\\
g^{t \psi} &=\frac{(\alpha\mu - q^2)b + a \alpha q}{\rho^2 \Delta},\\
g^{\phi \psi} &=-\frac{ab\mu + (a^2+b^2)q}{\Delta \rho^2},
\label{eqcinversemetricelements}
\end{aligned}
\end{align}

where

\begin{align}
\begin{aligned}
\alpha &=x+a^2,\\
\beta &= x+b^2.
\end{aligned} 
\end{align} 

 Inserting Eq. \eqref{eq:Sansatz} and Eq. \eqref{eqcinversemetricelements} into Eq. \eqref{eq:HamJac} yields a partial differential equation for  $S_r(r)$ and $S_\theta(\theta)$. This equation is separable in $r$ and $\theta$ and therefore both sides equate a separation constant $K$. We will fefer to $K$ as the Carter constant, as introduced in the Kerr spacetime \cite{Carter:1968rr}. Solving for $S_r(r)$, $S_\theta(\theta)$ and differentiating the action $S$ with respect to the constants of motion $(\delta, K, E, \Phi, \Psi)$ yields the geodesic equations

\begin{align}
\label{eq:geodesicequationsx}
\dot x^2 &= X,\\
\label{eq:geodesicequationstheta}
\dot \theta^2 &=\Theta,\\
\label{eq:geodesicequationsphi}
\dot \phi &= \frac{\Phi}{\sin^2\theta}-\frac{1}{\Delta}\Big[\Big((a^2-b^2)\beta+\mu b^2+2abq\Big)\Phi \\ & \quad + \Big(\mu ab+(a^2+b^2)q\Big)\Psi+\Big(a\beta\mu+b\beta q-aq^2\Big)E\Big],\\
\label{eq:geodesicequationspsi}
\dot \psi &=\frac{\Psi}{\cos^2\theta}-\frac{1}{\Delta}\Big[\Big(-(a^2-b^2)\alpha+\mu a^2+2abq\Big)\Psi\\ & \quad+ \Big(\alpha b\mu+a\alpha q-bq^2\Big)E+\Big(\mu ab+(a^2+b^2)q\Big)\Phi\Big],\\
\label{eq:geodesicequationst}
\dot t &=E\rho^2+\frac{1}{\Delta}\Big[ \Big(\mu\alpha\beta-q^2(\alpha+b^2\Big)E\\ & \quad+ \Big(a\beta\mu+b\beta q-aq^2\Big)\Phi+\Big(\alpha b \mu+a\alpha q-bq^2\Big)\Psi \Big],
\end{align} 

where the dot denotes the derivative with respect to a new parameter $\tau$, called Mino time \cite{Mino:2003yg}, related to $\lambda$ via

\begin{align}
d \tau = \frac{d\lambda}{\rho^2}
\end{align} 

as well as

\begin{align}
\begin{aligned}
 X&=4 \Big( \left(E^2 - \delta \right) \Delta x - \Delta K + \mathcal{E} + \mu\mathcal{M} + 2q\mathcal{Q} - q^2 \mathcal{P}\Big),  \\
\mathcal{E} &=  (a^2-b^2)(\beta \Phi^2 -\alpha \Psi^2), \\
\mathcal{M} &= \alpha\beta E^2 + 2 a \beta E \Phi + 2 \alpha b E \Psi + (b \Phi + a\Psi)^2, \\
\mathcal{Q} &=ab \left(\Phi^2 + \Psi^2\right) +  (a^2+b^2) \Phi \Psi + Eab \left(\frac{\Phi \beta}{a} + \frac{\Psi \alpha}{b} \right), \\
\mathcal{P} &=2a E\Phi + 2b E \Psi + (\alpha+b^2)E^2,\\
\Theta &= (E^2-\delta) \left(a^2 \cos^2 \theta + b^2 \sin^2 \theta \right) + K - \frac{\Phi^2}{\sin^2\theta}\\
& \quad - \frac{\Psi^2}{\cos^2 \theta}
\end{aligned} 
\end{align}

have been introduced for brevity.

\section{Discussion of the motion}

Before solving the geodesic equations, we will study several properties of the test particle motion by investigating the radial \eqref{eq:geodesicequationsx} and polar equation \eqref{eq:geodesicequationstheta}.

\subsection{$\theta$-motion}

The $\theta$-motion is described by Eq. \eqref{eq:geodesicequationstheta}. This geodesic equation is identical to the uncharged Myers-Perry case \cite{Diemer:2014lba}, i.e. the charge does not affect the $\theta$-motion of a neutral test particle. Furthermore, the $\theta$-motion is independent of the black hole's mass parameter $\mu$. As in the uncharged case, $\theta=0$ and $\theta=\frac{\pi}{2}$ can be reached only when $\Phi=0$ and $\Psi=0$, respectively. In both cases, the Carter constant $K$ can be expressed in terms of the remaining parameters $E$, $\Phi$, $\Psi$, $a$, $b$ and $\delta$. For $\theta = 0$ ($\Phi=0$), Eq. \eqref{eq:geodesicequationstheta} yields

\begin{align}
K = \Psi^2 - \left(E^2 - \delta \right) a^2
\end{align}

as well as for For $\theta = \frac{\pi}{2}$ ($\Psi=0$)

\begin{align}
K = \Phi^2 - \left(E^2 - \delta \right) b^2.
\label{eq:Kthetahalfpi}
\end{align}

A constant $\theta$-motion for some other $\theta_0 \in \left(0,\frac{\pi}{2}\right)$ requires

\begin{align}
\Theta(\theta_0) =0, \qquad \left.\frac{d\Theta}{d\theta}\right \vert_{\theta_0} =0.
\end{align}

To make our analysis simpler and our equations single-valued, we use $\xi = \cos^2 \theta$. So, $\xi \in [0,1]$ and our $\theta$ equation now becomes

\begin{align}
\dot\xi^2=a_3\xi^3+a_2\xi^2+a_1\xi+a_0 =: \Xi,
\end{align}

where the coefficients of the polynomial are given by

\begin{align}
\begin{aligned}
a_3 &=-4(E^2-\delta)(a^2-b^2),\\
a_2 &=4(E^2-\delta)(a^2-2b^2)-4K,\\
a_1 &=4(E^2-\delta)b^2-4\Phi^2+4\Psi^2+4K,\\
a_0 &=-4\Psi^2.
\end{aligned} 
\end{align}

We can now rewrite $\Xi$ as a function of the test particle's energy $E$

\begin{align}
\dot\xi^2 = \alpha_\xi \left(E^2 - V_{\theta}^2 \right),
\end{align}

where $V_{\theta}$ is the $\theta$-potential given by

\begin{align}
V_{\theta} = \sqrt{\frac{-\gamma_\xi}{\alpha_\xi}}
\end{align}

and $\alpha_\xi$, $\gamma_\xi$ are the coefficients of $\Xi=\alpha_\xi E^2+\gamma_\xi$. It is clear that the only regimes in which $\theta$-motion is possible are

\begin{enumerate}
\item[1)] $\alpha_\xi>0$ and $\vert E \vert > V_{\theta}$.
\item[2)] $\alpha_\xi<0$ and  $\vert E \vert < V_{\theta}$.
\end{enumerate}

\begin{figure*}[htbp]
\centering
\captionsetup{justification=justified}
\begin{minipage}[htbp]{0.33\textwidth}
   \includegraphics[width=0.93\textwidth]{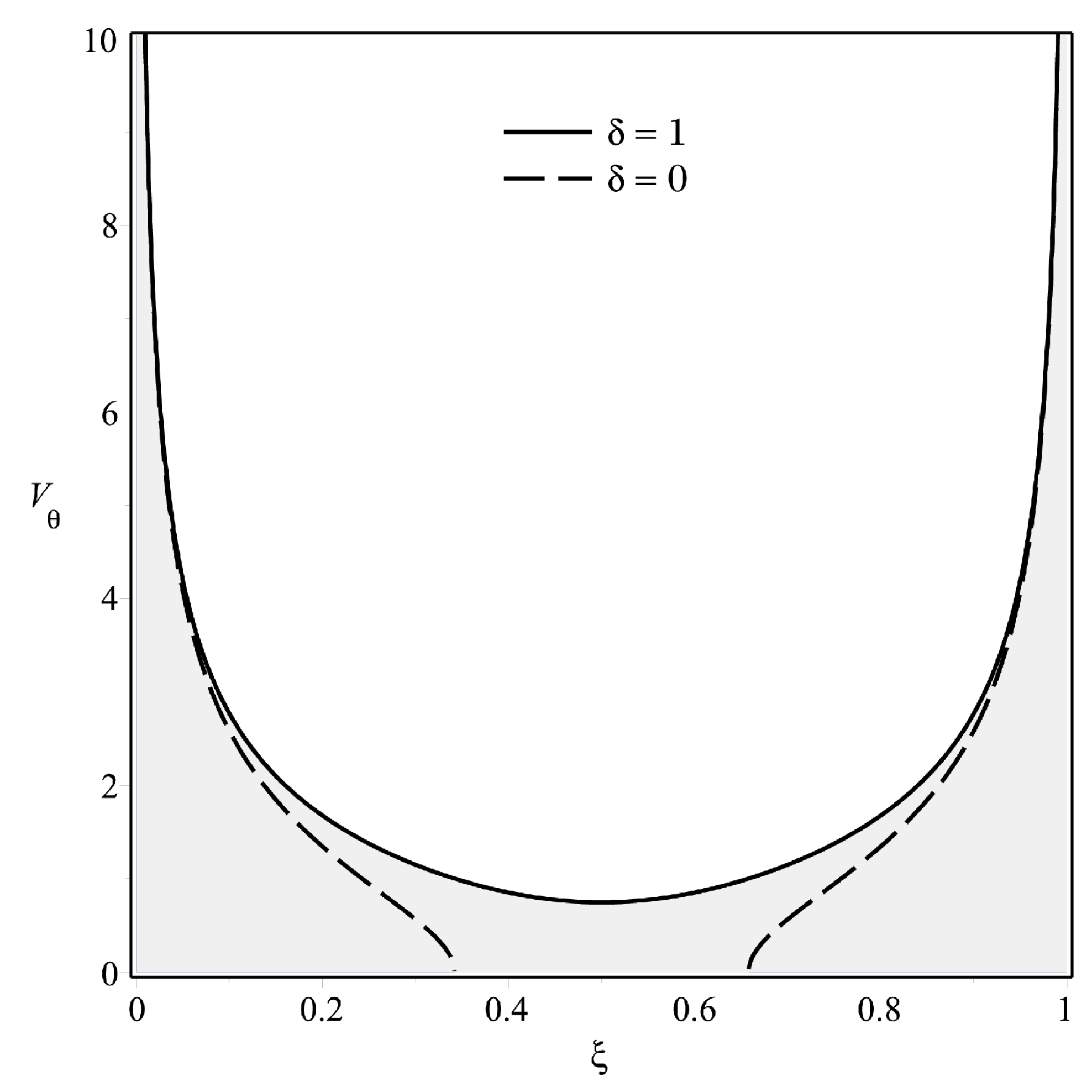}
\subcaption{$K = 0.4, \Phi=0.3, \Psi = 0.3,$\\  $a = 0.3, b=0.3$}
\label{fig:thetapota}
   \end{minipage}\hfill
   \begin{minipage}[htbp]{0.33\textwidth}
   \includegraphics[width=0.93\textwidth]{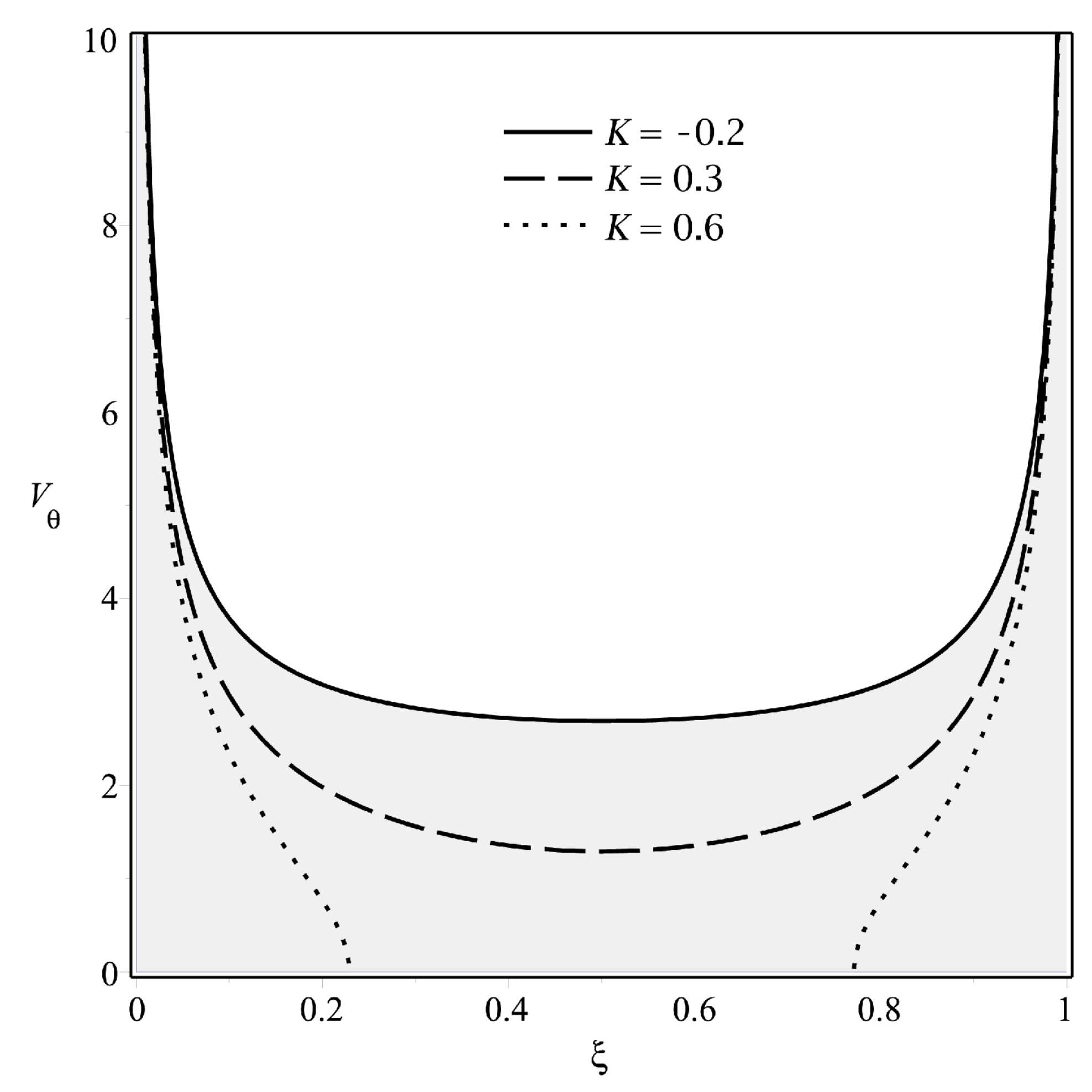}
\subcaption{$\Phi=0.3, \Psi = 0.3,  a = 0.3,$\\ $b=0.3, \delta = 1$}
\label{fig:thetapotb}
   \end{minipage}\hfill
   \begin{minipage}[htbp]{0.33\textwidth}
   \includegraphics[width=0.93\textwidth]{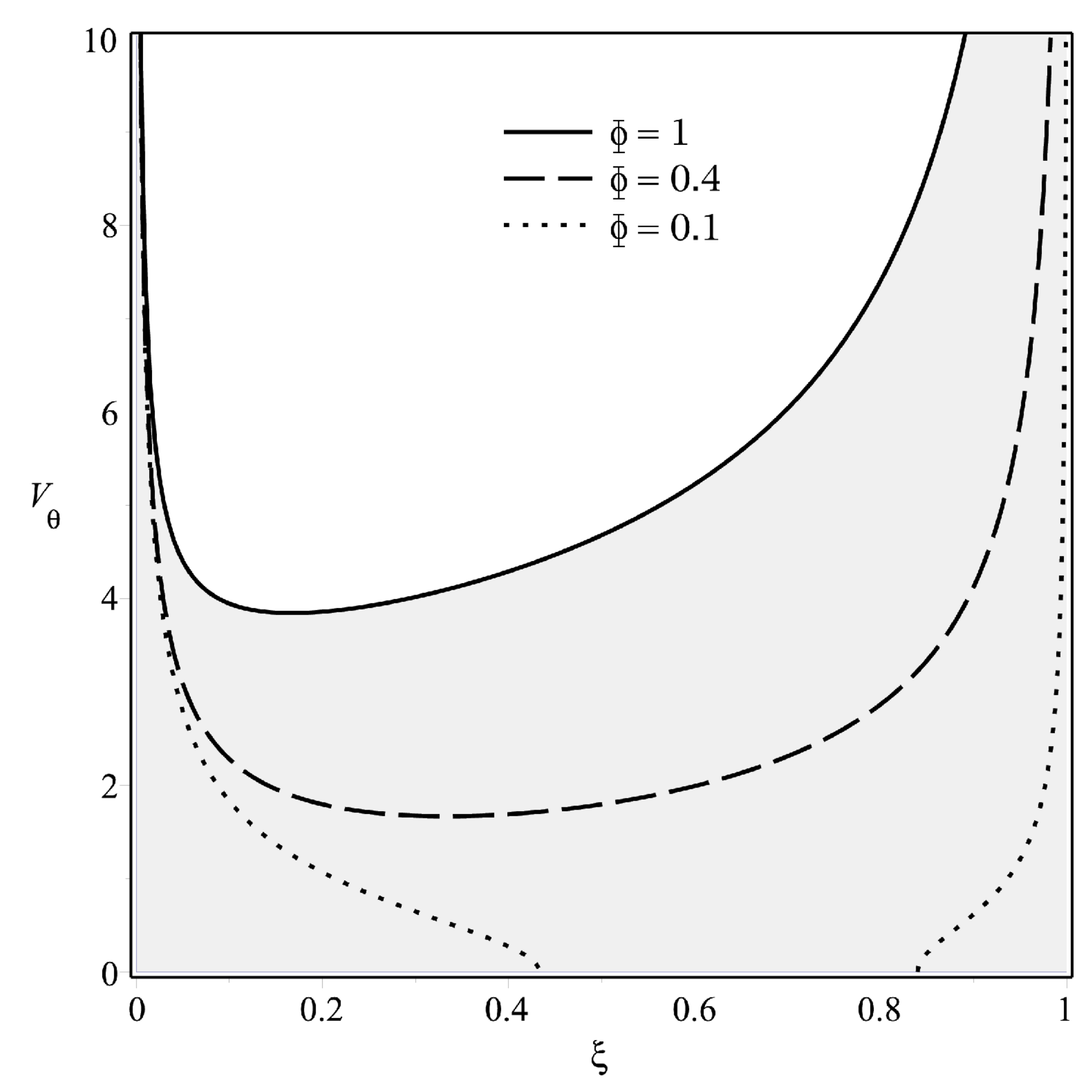}
\subcaption{$K = 0.2, \Psi = 0.2,  a = 0.3,$\\ $b=0.3, \delta = 1$}
\label{fig:thetapotc}
   \end{minipage}

\begin{minipage}[htbp]{0.33\textwidth}
   \includegraphics[width=0.93\textwidth]{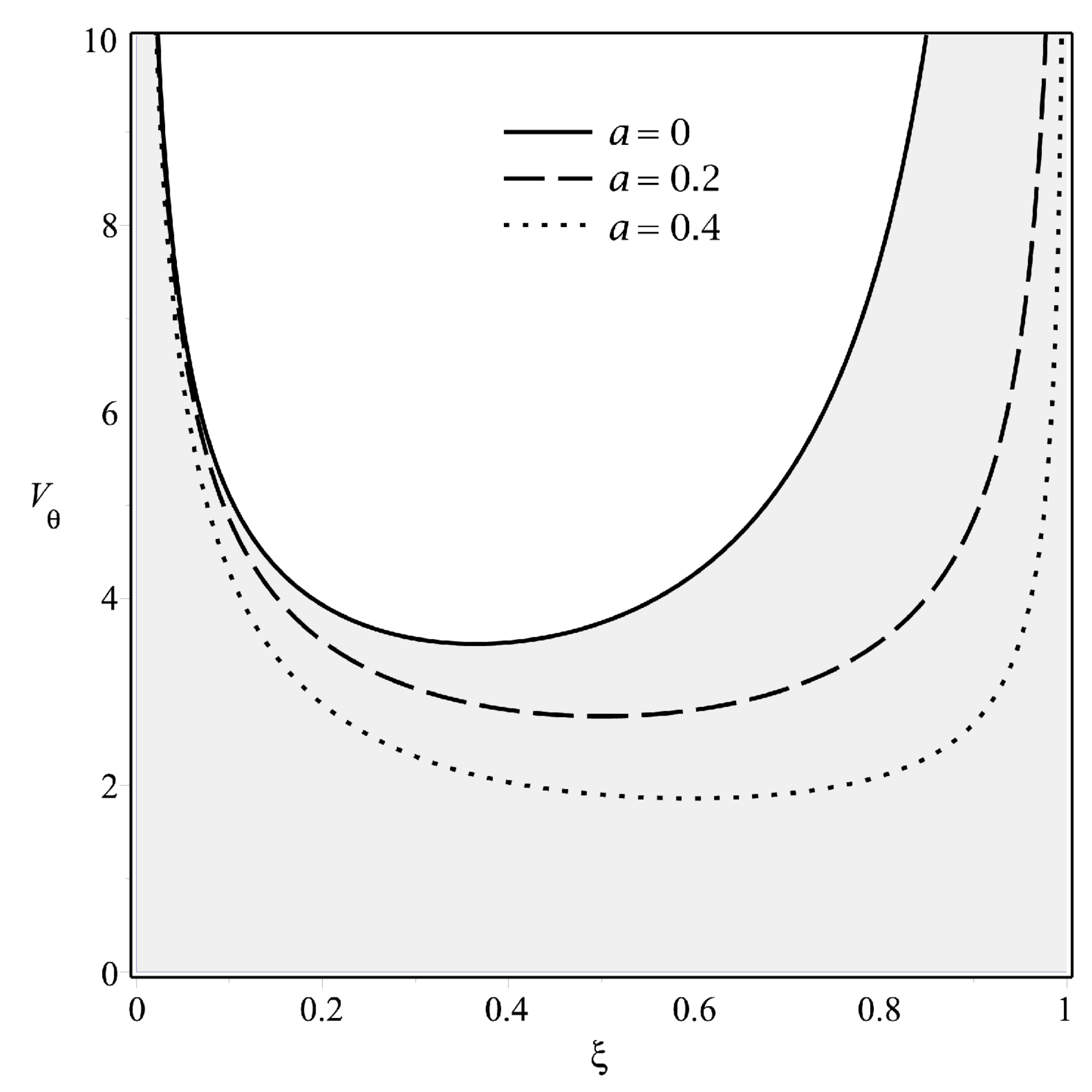}
\subcaption{$K = 0.1, \Phi=0.3, \Psi = 0.3,$\\ $b=0.2, \delta = 1$}
\label{fig:thetapotd}
   \end{minipage}\hfill
   \begin{minipage}[htbp]{0.33\textwidth}
   \includegraphics[width=0.93\textwidth]{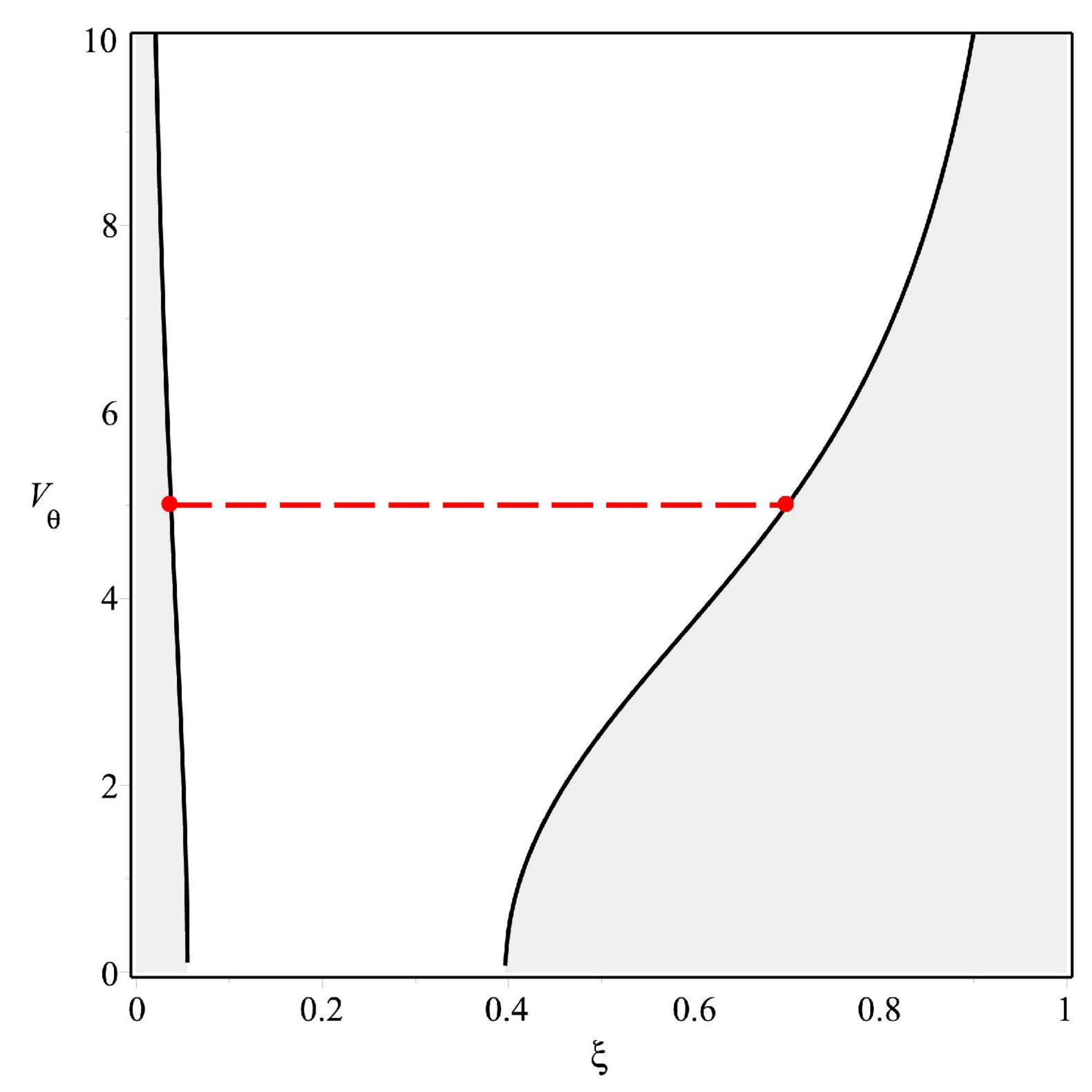}
\subcaption{$K = 1.8, \Phi=-1, \Psi = -0.2,$\\ $a = 0.3, b=0.1, \delta = 1$}
\label{fig:thetapote}
   \end{minipage}\hfill
   \begin{minipage}[htbp]{0.33\textwidth}
   \includegraphics[width=0.93\textwidth]{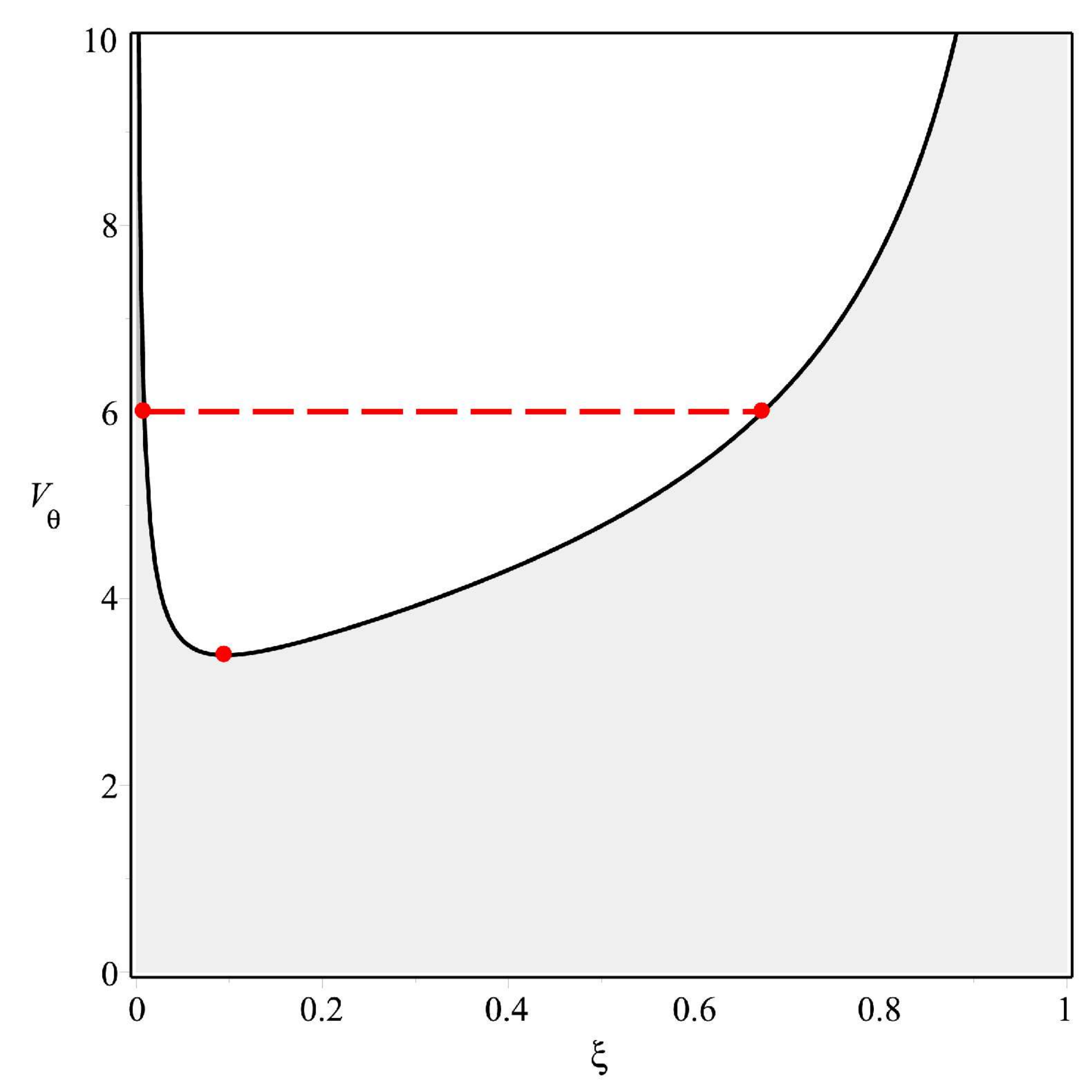}
\subcaption{$K = 1.7, \Phi=-1.39, \Psi = -0.1,$\\ $a = 0.4, b=0.2, \delta = 1$}
\label{fig:thetapotf}
   \end{minipage}
\captionsetup{justification=raggedright}
\caption{Plots of several $\theta$-potentials for $E \ge 0$. The grey-colored areas denote energy values, for which $\Xi<0$ and thus need to be excluded for physical test particle motion. Plots (a)-(d) emphasize the influence on the metric parameters $\delta, K, \Phi$ and $a$. Plots (e) and (f) show two typical potentials with possible $\theta$-motion (dashed red line) between two turning points (red points) or a constant $\theta$-motion related to a minimum in the potential.}
\label{fig:thetapot}
\end{figure*}

Since $\alpha_\xi$ is explicitely given by

\begin{align}
\alpha_\xi = 4\xi(1-\xi)\Big(\left(a^2-b^2\right)\xi+b^2\Big),
\end{align}

it is non-negative for $\xi \in [0,1]$. Therefore, we only need to consider regimes in which $\vert E \vert > V_{\theta}$. Due to the symmetry, it is sufficient to plot the $\theta$-potential for $E \ge 0$ (see Fig. \ref{fig:thetapot}).

Fig. \ref{fig:thetapota} represents a symmetric $\theta$-potential for both massive ($\delta=1$) and massless ($\delta=0$) test particles. The grey-colored area reprpesents energy values for which $\Xi<0$ and therefore no physical $\theta$-motion is possible. Due to Eq. \eqref{eq:geodesicequationstheta}, the difference between massive and massless test particle $\theta$-motion is only given by an energy shift. In the case of the parameters in Fig. \ref{fig:thetapota}, the massless test particle cannot reach energy values below the minimum of the $\theta$-potential, which vanishes for the massless test particle, allowing it to follow geodesics for arbitrary energy values.\\
 The same behavior occurs for the $K$-dependence in Fig. \ref{fig:thetapotb}.\\
Fig. \ref{fig:thetapotc} illustrates the $\Phi$-dependence. While obtaining higher values, the angular momentum $\Phi$ forces the test particle to be bounded between smaller values of $\xi$ (the potential bends towards $\xi = 0$ or $\theta = \frac{\pi}{2}$, respectively).\\
On the contrary, the potential bends towards $\xi = 1$ ($\theta =0$), when obtaining higher values of the black hole's rotation parameter $a$, as shown in Fig. \ref{fig:thetapotd}. Since $\Theta$ is invariant under the simultaneous transformations

\begin{align}
a \leftrightarrow b, \qquad \Phi \leftrightarrow \Psi \qquad \text{and} \qquad \theta \rightarrow \theta + \frac{\pi}{2},
\end{align}

the behaviour of the $\theta$-potential, when varying $\Psi$ and $b$, can be deduced from Fig. \ref{fig:thetapotc} and  Fig. \ref{fig:thetapotd}.\\
Fig. \ref{fig:thetapote} represents a typical $\theta$-potential, where the test particle may obtain arbitrary energy values, featuring a possible bounded $\theta$-motion.\\
Fig. \ref{fig:thetapotf} represents a typical $\theta$-potential, where the possible energy values are constrained by a lower boundary. In addition to a bounded $\theta$-motion, the test particle may now remain at a constant value of $\theta$ for this special energy at the potential minimum.

\subsection{$x$-motion}

The $x$-motion is described by Eq. \eqref{eq:geodesicequationsx}. Similiar to the $\theta$-equation, we can rewrite the $x$-equation via

\begin{align}
\dot x^2 = \alpha_x E^2 + \beta_x E + \gamma_x
\end{align}

and an effective potential, defined by

\begin{align}
\dot x^2 = \alpha_x \left(E-V_{\rm eff}^+\right) \left(E-V_{\rm eff}^- \right),
\label{eq:effpot}
\end{align}

can be expressed as

\begin{align}
V_{\rm eff}^\pm := \frac{-\beta_x \pm \sqrt{\beta_x^2 - 4\alpha_x \gamma_x}}{2 \alpha_x}.
\label{eq:Veff}
\end{align}

\begin{figure*}[htbp]
\centering
\captionsetup{justification=justified}
\begin{minipage}[htbp]{0.33\textwidth}
   \includegraphics[width=0.93\textwidth]{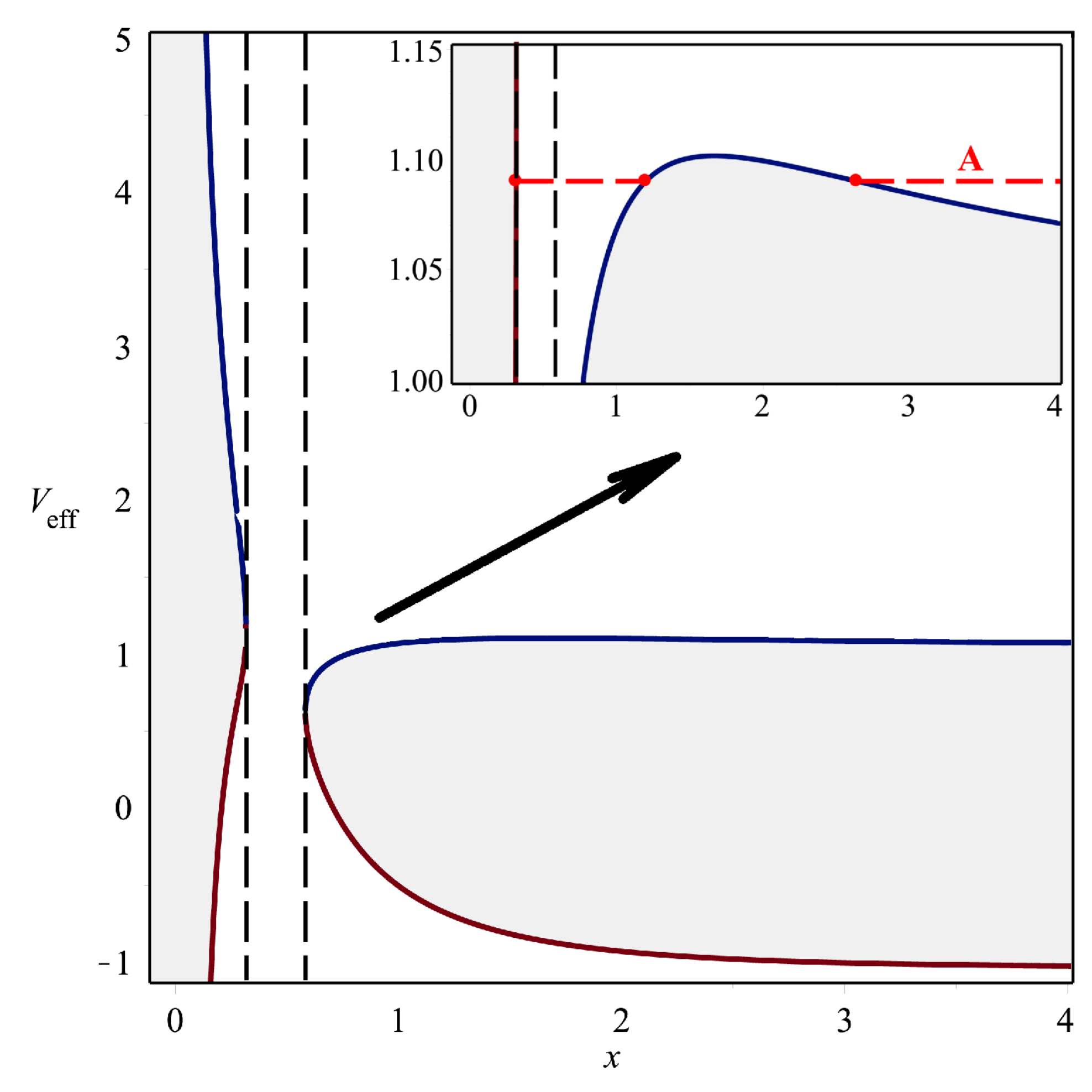}
\subcaption{$K = 1.8, \Phi=-1, \Psi = -0.2,$\\ $a = 0.3, b=0.1, q=0.4$}
\label{fig:xpota}
   \end{minipage}\hfill
   \begin{minipage}[htbp]{0.33\textwidth}
   \includegraphics[width=0.93\textwidth]{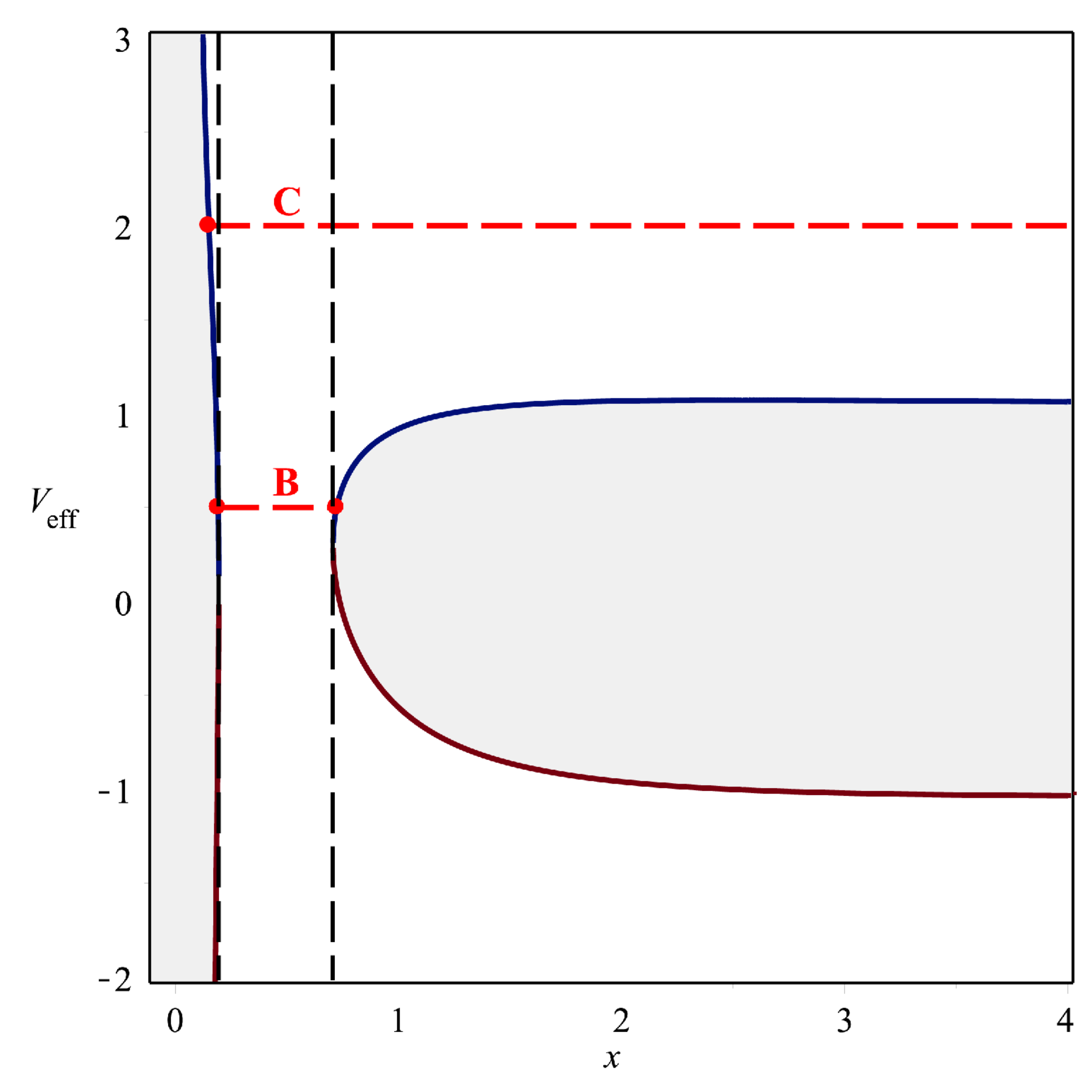}
\subcaption{$K = 1.8, \Phi=-1, \Psi = -0.2,$\\ $a = 0.3, b=0.1, q=-0.4$}
\label{fig:xpotb}
   \end{minipage}\hfill
   \begin{minipage}[htbp]{0.33\textwidth}
   \includegraphics[width=0.93\textwidth]{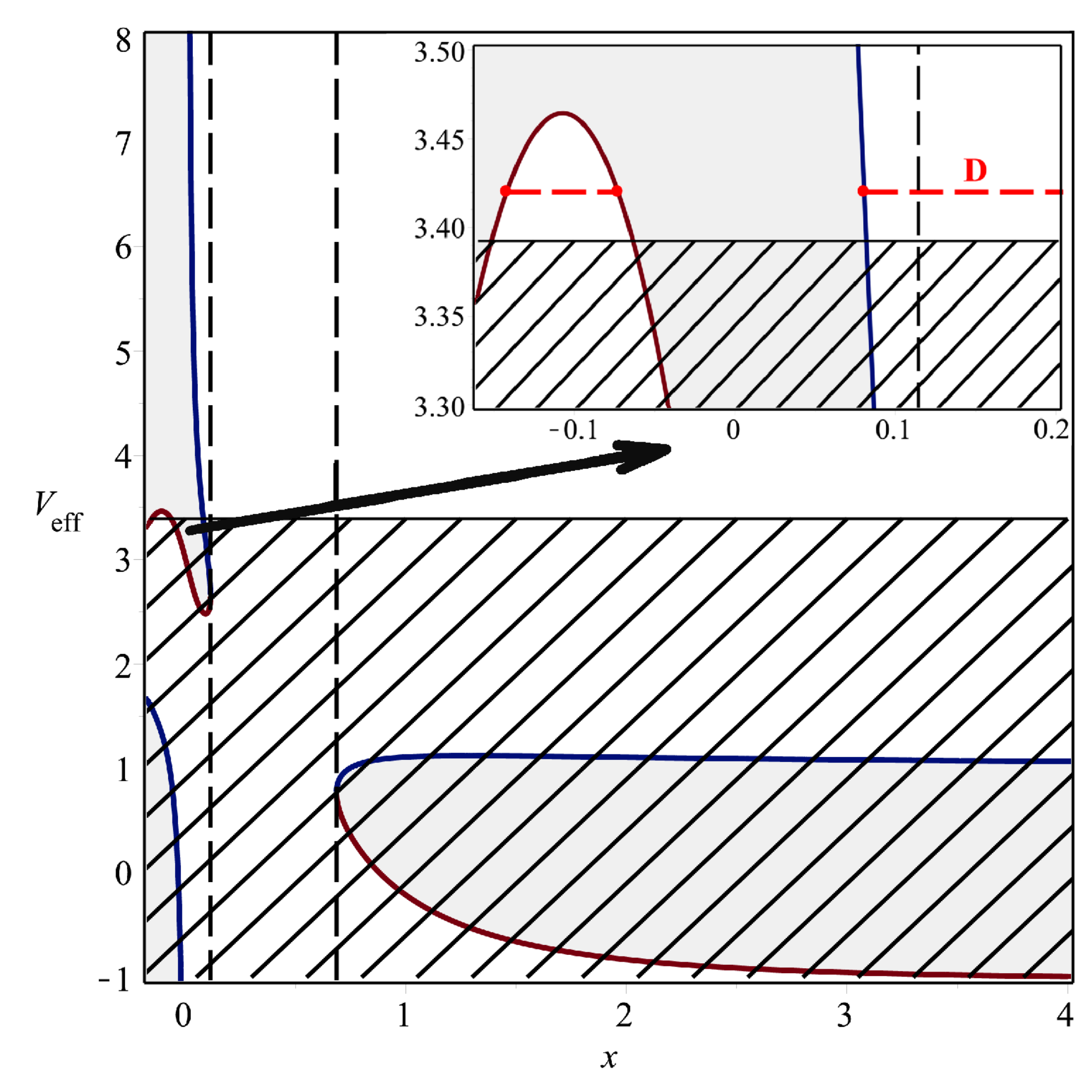}
\subcaption{$K = 1.7, \Phi=-1.39, \Psi = -0.1, $\\ $a = 0.4, b=0.2, q = 0.2$}
\label{fig:xpotc}
   \end{minipage}

\begin{minipage}[htbp]{0.33\textwidth}
   \includegraphics[width=0.93\textwidth]{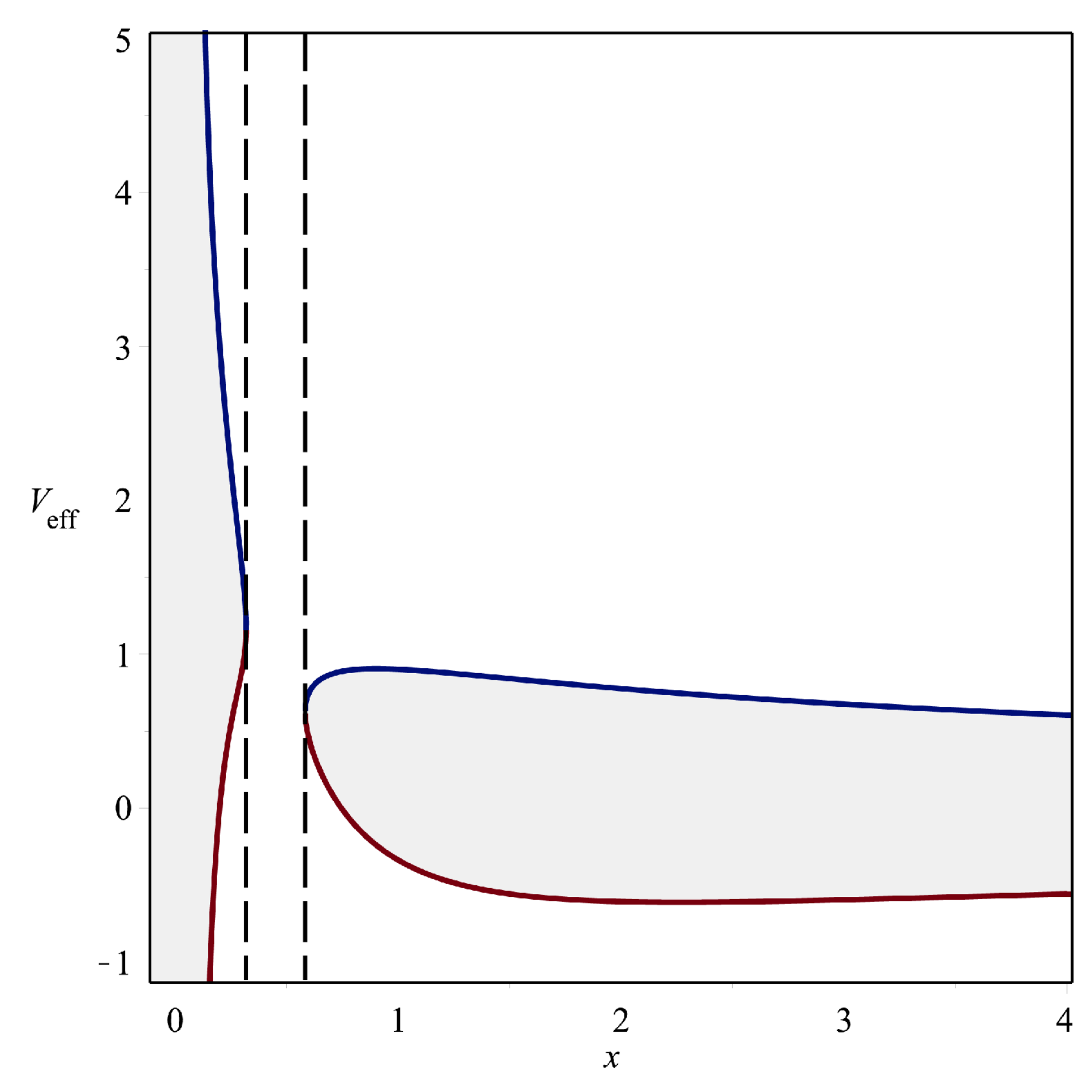}
\subcaption{$K = 1.8, \Phi=-1, \Psi = -0.2,$\\ $a = 0.3, b=0.1, q=0.4$}
\label{fig:xpotd}
   \end{minipage}\hfill
   \begin{minipage}[htbp]{0.33\textwidth}
   \includegraphics[width=0.93\textwidth]{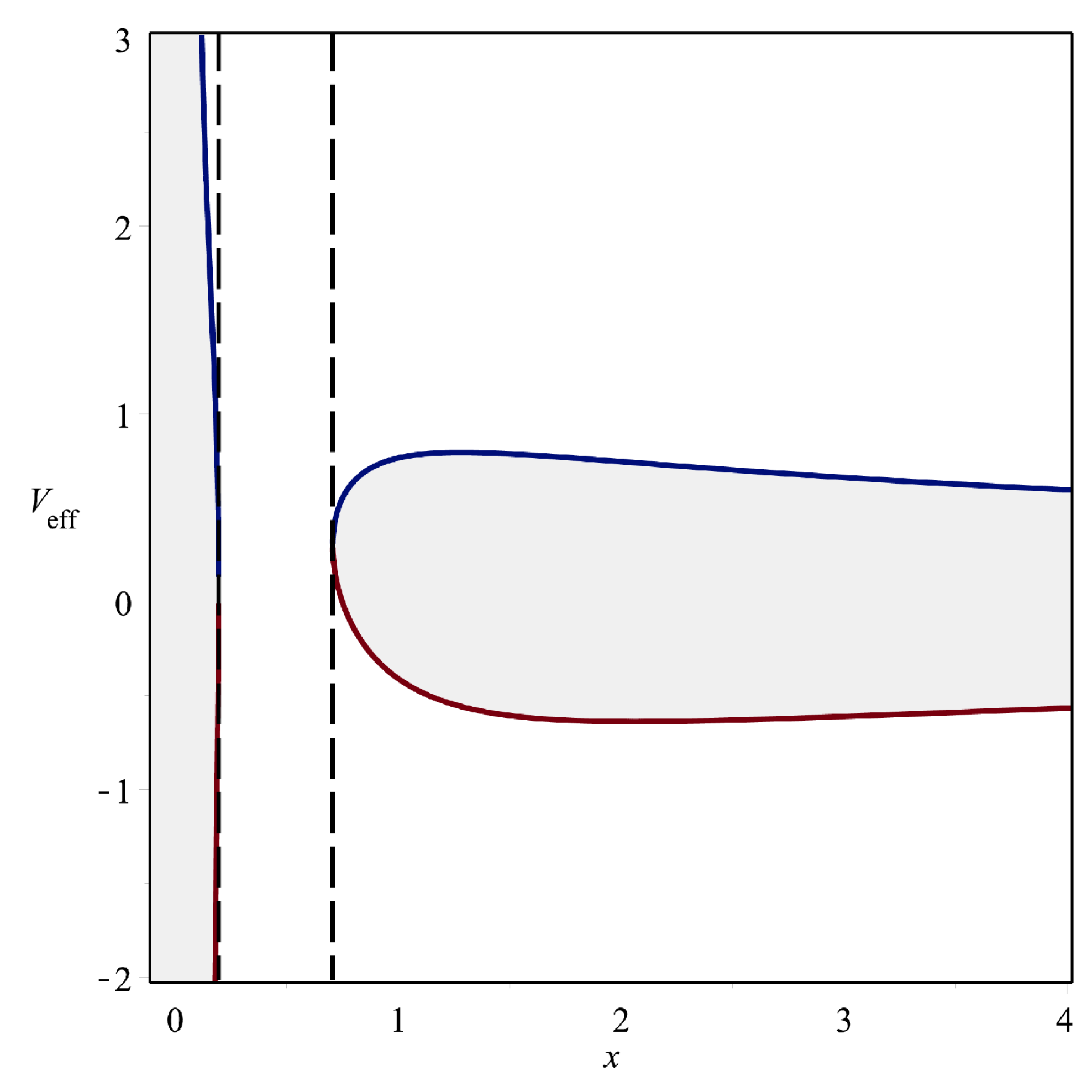}
\subcaption{$K = 1.8, \Phi=-1, \Psi = -0.2,$\\ $a = 0.3, b=0.1, q=-0.4$}
\label{fig:xpote}
   \end{minipage}\hfill
   \begin{minipage}[htbp]{0.33\textwidth}
   \includegraphics[width=0.93\textwidth]{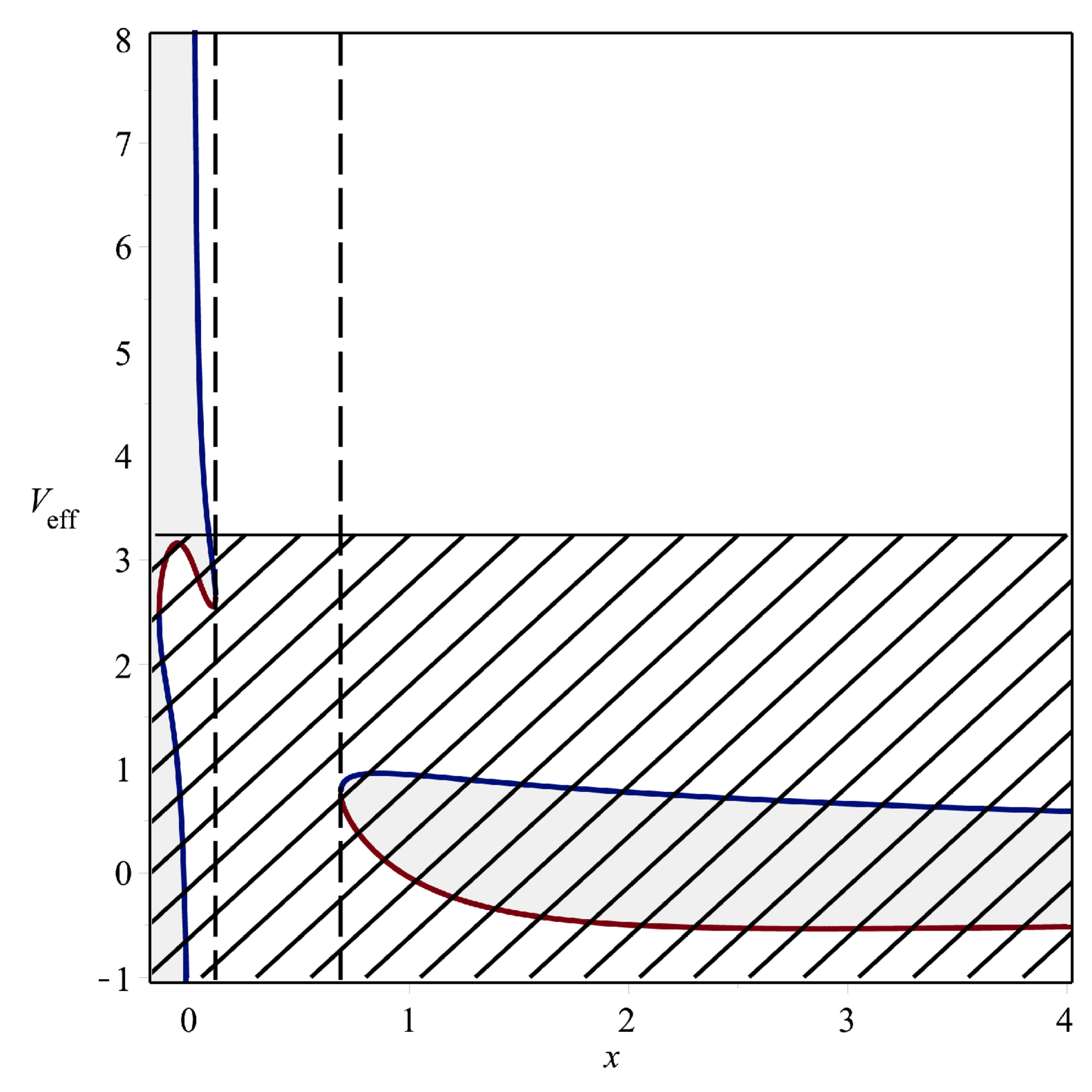}
\subcaption{$K = 1.7, \Phi=-1.39, \Psi = -0.1, $\\ $a = 0.4, b=0.2, q = 0.2$}
\label{fig:xpotf}
   \end{minipage}
\captionsetup{justification=raggedright}
\caption{Effective potentials for massive (a-c) and massless (d-f) test particles. The grey are are physically forbidden by the radial geodesic equation. Regions which are energetically restricted due to the conditions of the $\theta$-polynomial are hatched. Different orbit types (A-D) are given by the red lines of constant energy for massive test particle and the corresponding massless potential is shown below.}
\label{fig:xpot}
\end{figure*}

The condition $\dot x^2 = 0$ defines the radial turning points for some value of the test particle's energy $E$. Thus, the possible orbit types of the test particle motion can be illustrated by this effective potentials (see Fig. \ref{fig:xpot}). Parameter values for which the right-hand side of Eq. \eqref{eq:effpot} becomes negative are physically not allowed. Additionally, we have to consider the results of the $\theta$-potentials. 

By the means of these effective potentials we conclude that there are the following different types of orbits:

\begin{itemize}
	\item[] {\bf Escape orbit (EO):} $x$ starts from infinity and approaches a periapsis and goes back to infinity.\\[-15pt]
	\item[] {\bf Two-world escape orbit (TEO):} A special case of an escape orbit, where the radial turning point lies behind both horizons. Due to causality, it cannot repass both horizons to the former universe but to a different universe.\\[-15pt]
	\item[] {\bf Bound orbit (BO):} $x$ oscillates between two radial turning points $x_1, x_2$, where either $x_1, x_2 \le x_-$ or $x_1, x_2 \ge x_+$.\\[-15pt]
	\item[] {\bf Many-world bound orbit (MBO):} A special case of a bound orbit, where $x_1 \le x_-$ and $x_2 \ge x_+$. For the same reasons mentioned considering the TEO, each time both horizons are passed through, the former universe cannot be reentered. So after every oscillation, the test particle enters a different universe.\\[-15pt]
	\item[] {\bf Terminating orbit (TO):} $x$ starts from infinity and hits the curvature singularity at $x = x_{\rm s}$.\\[-10pt]
\end{itemize}

Orbit type A represents either a many-world bound orbit or an escape orbit depending on the initial radial position of the test particle. Orbit types B and C describe many-world bound orbits or two-world escape orbits, respectively. Orbit type D denotes a bound orbit or a two-world escape orbit again depending on the initial radial position of the test particle. \\

Obviously, the effective potentials for massive and massless test particles are quite similar. Anyway, Figs. \ref{fig:xpot} reveal that, for an equal set of parameters $(K,\Phi,\Psi,a,b,q,\mu)$, there might be orbits, which are only valid for one type of test particle. While there is a possible bound orbit in Fig. \ref{fig:xpotc}, it obviously vanishes in Fig. \ref{fig:xpotf}. Nevertheless, the same types of orbits are possible for either massive and massless test particles for arbitrary sets of parameters.\\
In Fig. \ref{fig:xpota} and Fig. \ref{fig:xpotb} the sign of the black hole charge $q$ changes, while the remaining parameters are retained. As described in Fig. \eqref{fig:horizons}, the horizons do not coincide in both cases as well as the effective potentials differ, especially for $x < x_-$. This is again an effect of the Chern-Simons term.\\
Fig. \ref{fig:xpotc} and Fig. \ref{fig:xpotf} also include a restricted region due to the $\theta$-potential. In some cases, test particle orbits with energy equal to this $\theta$-boundary are known to allow terminating orbits in the five-dimensional Myers-Perry spacetime (see \cite{Kagramanova:2012hw,Diemer:2014lba}). The same is valid for the five-dimensional, charged, rotating, EMCS spacetime.\\

In Tab. \ref{tab:orbittypes} all possible types of orbits are summarized:

\begin{table}[H]
	\centering
\resizebox{0.45\textwidth}{!}{%
	\begin{tabular}{|c|c|c|c|c|}
		\hline
	Orbit type & Orbits & Radial zeros & Range of $x$\\ \hline \hline
		A & MBO, EO  & 3 & \mboeo \\ \hline
		B & MBO  & 2 & \teo \\ \hline
		C & TEO & 1 & \mbo \\ \hline
		D & BO, TEO & 3 & \boteo \\ \hline
		E & TO  & 0 & \tos \\ \hline
	\end{tabular}%
	}
	\caption{Summary of the possible orbit types in the five-dimensional  charged rotating EMCS spacetime. Thick horizontal lines represent possible orbits, radial turning points are illustrated by big dots and the horizons as well as the singularity are shown as two thin vertical lines or one thick vertical line, respectively.}
	\label{tab:orbittypes}
\end{table}

\section{Analytic solutions}

\subsection{$\theta$-equation}

As stated earlier, we obtain the following differential equation for $\xi$ 

\begin{align}
\dot\xi^2=a_3\xi^3+a_2\xi^2+a_1\xi+a_0,
\label{eq:xieq}
\end{align}

so that the right side of the equation is a cubic polynomial in $\xi$. We follow the same method as in the uncharged Myers-Perry spacetime \cite{Diemer:2014lba}. We make the coordinate transformation $\xi=\frac{1}{a_3}\left(4y-\frac{a_3}{3}\right)$ and obtain

\begin{align}
\dot y^2=4y^3-g_2y-g_3,
\end{align}

where the coefficients in the polynomial are called Weierstrass invariants and are given by

\begin{align}
\begin{aligned}
g_2 &=\frac{a_2^2}{12}-\frac{a_1a_3}{4},\\
g_3 &=\frac{a_1a_2a_3}{48}-\frac{a_0a_3^3}{16}-\frac{a_2^3}{216}.
\end{aligned}
\end{align}

Following the standard procedure for this elliptic integral, we get 

\begin{align}
y(\tau) &=\wp(\tau-\tau_{\rm in}^{\theta}),
\end{align}

where $\wp(\tau)$ denotes the Weierstrass elliptic function and

\begin{align}
\tau^\theta_{\rm in} &=\tau_{\rm in}+\int_{y_{\rm in}}^\infty\frac{dy}{\sqrt{4y^3-g_2y-g_3}}.
\end{align}

Furthermore, $\tau_{\rm in}$ denotes initial value and the second term in the expression of $\tau^\theta_{\rm in}$ is a constant as it is a definite integral between two fixed points. Thus, it can be evaluated in terms of the parameter values. To bring our solution into terms of $\theta$ instead of $y$ we substitute back to $\xi$ and then $\theta$

\begin{align}
\theta(\tau)=\arccos \left(\frac{1}{a_3}\left(4\wp\left(\tau-\tau^{\theta}_{\rm in};g_2,g_3\right)-\frac{a_2}{3}\right) \right).
\end{align}

\subsection{$x$-equation}

Proceeding as above, we first write the differential equation for the $x$-coordinate as a polynomial in $x$

\begin{align}
\dot x^2=16\Delta^2\chi=b_3x^3+b_2x^2+b_1x+b_0.
\label{eq:xeq}
\end{align}

Since this is of the same form as the $\theta$-equation, substituting $x=\frac{1}{b_3}\left(4z-\frac{b_3}{3}\right)$ will give us the Weierstrass equation with constants $h_2$ and $h_3$. From this, we find the solution to be 

\begin{align}
x(\tau)=\frac{1}{b_3}\left(4\wp\left(\tau-\tau^{x}_{\rm in};h_2,h_3\right)-\frac{b_3}{3}\right),
\end{align}

where as before, $\tau^{x}_{\rm in}$ is a constant intial value given by

\begin{align}
\tau^x_{\rm in}=\tau_{\rm in}+\int_{z_{\rm in}}^\infty\frac{dz}{\sqrt{4z^3-h_2z-h_3}}.
\end{align}

\subsection{$\phi$-equation}

The differential equation for $\phi$ involves dependencies on $x$ and $\theta$. We can deal with both of these parts separately

\begin{align}
\begin{aligned}
d\phi_\theta &=\frac{\Phi}{\sin^2\theta}d\tau=\frac{1}{1-\xi}\frac{d\xi}{\sqrt{\Xi}}=R^\phi(y)\frac{dy}{\sqrt{Y}},\\
d\phi_x &=-\frac{1}{\Delta}\Big[\Big((a^2-b^2)\beta+\mu b^2+2abq\Big)\Phi \\ 
&\quad \,+ \Big(\mu ab+(a^2+b^2)q\Big)\Psi+\Big(a\beta\mu+b\beta q-aq^2\Big)E\Big]\\
&=\tilde{R}^\phi(y)\frac{dz}{\sqrt{Z}}.
 \label{eq:phiparts}
\end{aligned}
\end{align}

In the first step, we change the differential from $d\tau$ to $d\xi$ or $dx$ via Eq. \eqref{eq:xieq} and Eq. \eqref{eq:xieq}, respectively, and then we make the substitutions used  earlier to bring the equations to the Weierstrass form. Here, $R^\phi$ and $\tilde{R}^\phi$ are rational functions that can be expressed via partial fraction decomposition as

\begin{align}
\begin{aligned}
R^\phi(y) &=\frac{G^\phi}{y-p_1},\\
\tilde{R}^\phi(z) &=\frac{H_1^\phi}{z-q_1}+\frac{H_2^\phi}{z-q_2},
\end{aligned}
\end{align}

where $p_1,q_1,q_2,G^\phi,H_1^\phi,H_2^\phi$ are the decomposition constants. Thus, we can integrate the differentials to get

\begin{align}
\begin{aligned}
\phi_\theta-\phi^0_\theta &=\int_{y_{in}}^y \frac{G^\phi}{y-p_1} \frac{dy}{\sqrt{Y}} = \int_{v_{in}}^v \frac{G^\phi}{\wp(v)-\wp(v_1)} dv,\\
\phi_x-\phi^0_x &=\int_{z_{in}}^z \frac{H_1^\phi}{z-q_1}+\frac{H_2^\phi}{z-q_2} \frac{dz}{\sqrt{Z}}\\ 
&= \int_{w_{in}}^w \frac{H_1^\phi}{\wp(w)-\wp(w_1)}+\frac{H_2^\phi}{\wp(w)-\wp(w_2)} dw,
\end{aligned}
\end{align}

where we substituted $y = \wp(v)$ as well as $z = \wp(w)$ anddefined $v_1, w_1$ and $w_2$ as $\wp(v_1)=p_1$, $\wp(w_1)=q_1$ and $\wp(w_2)=q_2$. Further, we introduced the corresponding initial values $\phi_{\rm in}^\theta$ and $\phi_{\rm in}^x$. Using the identity

\begin{align}
\frac{\wp^\prime(v)}{\wp(w)-\wp(v)}=\zeta(w-v)+\zeta(w+v)+2\zeta(v)
\end{align}

and the fact that the Weierstrass $\zeta$-functions are in turn the total derivatives of the logarithm of the Weierstrass-$\sigma$ functions, we get

\begin{align}
\begin{aligned}
\phi_\theta &=\frac{G^\phi}{\wp^\prime(v_1)}\bigg[\ln \left(\frac{\sigma(v(\tau)-v_1)}{\sigma(v(\tau)+v_1)}\right) - \ln \left(\frac{\sigma(v_{in}-v_1)}{\sigma(v_{in}+v_1)}\right)\\
& \quad \,+2\zeta(v_1)(v(\tau)-v_1)\bigg]+\phi_{\rm in}^\theta,\\
\phi_x &=\frac{H_1^\phi}{\wp^\prime(w_1)}\bigg[ \ln \left(\frac{\sigma(w(\tau)-w_1)}{\sigma(w(\tau)+w_1)}\right) - \ln \left(\frac{\sigma(w_{in}-w_1)}{\sigma(w_{in}+w_1)}\right)\\ 
&\quad \,+2\zeta(w_1)(w(\tau)-w_1)\bigg]\\
& \quad \,+ \frac{H_2^\phi}{\wp^\prime(w_2)}\bigg[ \ln \left(\frac{\sigma(w(\tau)-w_2)}{\sigma(w(\tau)+w_2)}\right)\\
&\quad \, - \ln \left(\frac{\sigma(w_{in}-w_2)}{\sigma(w_{in}+w_2)}\right)+2\zeta(w_2)(w(\tau)-w_2)\bigg]+\phi_{\rm in}^x.
\end{aligned}
\end{align}

\subsection{$\psi$-equation}

Analogous to the solution of the $\phi$-equation, the $\psi$-equation depends on both $x$ and $\theta$

\begin{align}
\begin{aligned}
d\psi_\theta &=R^\psi(y)\frac{dy}{\sqrt{Y}},\\
d\psi_x &=\tilde{R}^\psi(z)\frac{dz}{\sqrt{Z}}.
 \label{eq:psiparts}
\end{aligned}
\end{align}

Since the equations are symmetric under the following transformations

\begin{align}
\Phi \leftrightarrow \Psi,\quad a \leftrightarrow b,\quad \theta(\tau) \leftrightarrow \theta(\tau)+\frac{\pi}{2},
\end{align}

we can use them on the $\phi$-equation to obtain the $\psi$-equation. We get

\begin{align}
\begin{aligned}
\psi_\theta &=\frac{G^\psi}{\wp^\prime(v_2)}\bigg[ \ln \left(\frac{\sigma(v(\tau)-v_2)}{\sigma(v(\tau)+v_2)}\right) - \ln \left(\frac{\sigma(v_{in}-v_2)}{\sigma(v_{in}+v_2)}\right)\\
& \quad \,+2\zeta(v_2)(v(\tau)-v_2)\bigg]+\psi_{\rm in}^\theta,\\
\psi_x &=\frac{H_1^\psi}{\wp^\prime(w_1)}\bigg[ \ln \left(\frac{\sigma(w(\tau)-w_1)}{\sigma(w(\tau)+w_1)}\right) - \ln \left(\frac{\sigma(w_{in}-w_1)}{\sigma(w_{in}+w_1)}\right)\\
&\quad \, +2\zeta(w_1)(w(\tau)-w_1)\bigg]\\
& \quad \, + \frac{H_2^\psi}{\wp^\prime(w_2)}\bigg[ \ln \left(\frac{\sigma(w(\tau)-w_2)}{\sigma(w(\tau)+w_2)}\right)\\
& \quad \, - \ln \left(\frac{\sigma(w_{in}-w_2)}{\sigma(w_{in}+w_2)}\right)+2\zeta(w_2)(w(\tau)-w_2)\bigg]+\psi_{\rm in}^x
\end{aligned}
\end{align}

with some initial values $\psi_{\rm in}^\theta$ and $\psi_{\rm in}^x$.
\subsection{$t$-equation}

Again, we split the $dt$-differential into an $x$- and a $\theta$-dependent part

\begin{align}
\begin{aligned}
dt_\theta &=E(a^2\cos^2\theta+b^2\sin^2\theta)d\tau = R^t(y)\frac{dy}{\sqrt{Y}},\\
dt_x &= \frac{1}{\Delta}\Big[ \Big(\mu\alpha\beta-q^2(\alpha+b^2\Big)E + \Big(a\beta\mu+b\beta q-aq^2\Big)\Phi\\
& \quad \, +\Big(\alpha b \mu+a\alpha q-bq^2\Big)\Psi \Big] d\tau\\
& =\tilde{R}^t(z)\frac{dz}{\sqrt{Z}}
\end{aligned}
 \label{eq:tparts}
\end{align}

with some rational functions $R^t$ and $\tilde{R}^t$. After substituting the form of $x(\tau)$ and $\theta(\tau)$ and integrating, we get

\begin{align}
\begin{aligned}
t_\theta &=-J_1^t\big(\zeta(v(\tau))-\zeta(v_{in})\big)-J_0^t \big(v(\tau)-v_{in}\big)+t_{\rm in}^\theta,\\
t_x &=-K_1^t\big(\zeta(w(\tau))-\zeta(w_{in})\big)-K_0^t\big(w(\tau)-w_{in}\big)\\
& \quad \,+\frac{H_1^t}{\wp'(w_1)}\bigg[ \ln \left(\frac{\sigma(w(\tau)-w_1)}{\sigma(w(\tau)+w_1)}\right) - \ln \left(\frac{\sigma(w_{in}-w_1)}{\sigma(w_{in}+w_1)}\right)\\
& \quad \,+2\zeta(w_1)(w(\tau)-w_1) + \frac{H_2^t}{\wp'(w_2)}\bigg[ \ln \left(\frac{\sigma(w(\tau)-w_2)}{\sigma(w(\tau)+w_2)}\right)\\
& \quad \, - \ln \left(\frac{\sigma(w_{in}-w_2)}{\sigma(w_{in}+w_2)}\bigg]+2\zeta(w_2)(w(\tau)-w_2)\right)+t_{\rm in}^x,
\end{aligned}
\end{align}

where $J_1^t, J_2^t, H_1^t$ and $H_2^t$ are again the decomposition constants, $t_{\rm in}^\theta$, $t_{\rm in}^x$ are inital values and the other quantities are defined as in the $\phi$-solution.

\section{Observables}

In this section we want to present the analytical expressions for some spacetime observables. These quantities are e.g. the light deflection for escape orbits, the perihelion shift for bound orbits or the Lense-Thirring effect. The expressions are similar to those that we calculated for the uncharged case of the five-dimensional Myers-Perry spacetime \cite{Diemer:2014lba}. Consequently, we follow along the lines of this paper and \cite{Hackmann:2010zz,Drasco:2003ky,Fujita:2009bp} for calculations.

\subsection{Deflection angle}

The deflection angle of an escape orbit with radial turning point $x_0 = \frac{1}{b_3} \left(4 z_0 - \frac{b_2}{3}\right)$ can be determined by calculating the values $\tau_\pm^\infty$ of the Mino time for which $x(\tau_\pm^\infty) = \infty$. This yields

\begin{align}
\tau_\pm^\infty - \tau_{\rm in} = \pm \int_{x_0}^\infty \frac{dx}{4 \Delta \sqrt{\mathcal{X}}} =  \pm \int_{z_0}^\infty\frac{dz}{\sqrt{4z^3-h_2z-h_3}},
\label{eq:deflection}
\end{align}

where the sign is related to both branches of the expressions, respectively. The total change of the angular coordinates may now be calculated as

\begin{align}
\begin{aligned}
\Delta \theta &= \theta \left(\tau_+^\infty \right) - \theta \left(- \tau_-^\infty \right),\\
\Delta \phi &= \phi \left(\tau_+^\infty \right) - \phi \left(- \tau_-^\infty \right),\\
\Delta \psi &= \psi \left(\tau_+^\infty \right) - \psi \left(- \tau_-^\infty \right),
\end{aligned}
\end{align}

so that the related deflection angles $\delta \theta$, $\delta \phi$ and $\delta \psi$ are defined as the total change of the angular coordinate minus $\pi$ \cite{Hartle:2003yu}.

\subsection{Perihelion shift and Lense-Thirring effect}

Now, we want to consider the perihelion shift and the Lense-Thirring effect for bound orbits or many-world bound orbits. The radial $x$- and polar $\theta$-motion are periodic with periods 

\begin{align}
\begin{aligned}
\omega_\theta &= 2 \int_{\theta_{\rm min}}^{\theta_{\rm max}} \frac{\mathrm d\theta}{\sqrt{\Theta}} = 2 \int_{e_1^y}^{e_2^y} \frac{\mathrm dy}{\sqrt{Y}} = 2 \omega_1^y,\\
\omega_x &= 2 \int_{x_{\rm min}}^{x_{\rm max}} \frac{\mathrm dx}{\sqrt{X}} = 2 \int_{e_1^z}^{e_2^z} \frac{\mathrm dz}{\sqrt{Z}} = 2 \omega_1^z,
\label{eq:periods}
\end{aligned}
\end{align}

which are related to the first fundamental period $\omega_1^{y,z}$ of the $\wp$-function.

The corresponding orbital frequencies with respect to the Mino time $\tau$ are $\Upsilon_\theta = \frac{2\pi}{\omega_\theta}$ and $\Upsilon_x = \frac{2\pi}{\omega_x}$. The orbital periods of the remaining coordinates are given by

\begin{align}
\begin{aligned}
\Upsilon_\phi &= \frac{2}{\omega_\theta} \int_{\theta_{\rm min}}^{\theta_{\rm max}} \mathrm d\phi_\theta + \frac{2}{\omega_x} \int_{x_{\rm min}}^{x_{\rm max}} \mathrm d\phi_x,\\
\Upsilon_\psi &= \frac{2}{\omega_\theta} \int_{\theta_{\rm min}}^{\theta_{\rm max}}\mathrm d\psi_\theta + \frac{2}{\omega_x} \int_{x_{\rm min}}^{x_{\rm max}} \mathrm d\psi_x,\\
\Gamma &= \frac{2}{\omega_\theta} \int_{\theta_{\rm min}}^{\theta_{\rm max}} \mathrm dt_\theta + \frac{2}{\omega_x} \int_{x_{\rm min}}^{x_{\rm max}} \mathrm dt_x,
\label{eq:frequencies}
\end{aligned}
\end{align}

where the corresponding differentials are given by Eq. \eqref{eq:phiparts}, Eq. \eqref{eq:psiparts} and Eq. \eqref{eq:tparts}. Finally, the orbital frequencies with respect to the coordinate time $t$ are given by

\begin{align}
\begin{aligned}
\Omega_\theta =  \frac{\Upsilon_\theta}{\Gamma}, \quad \Omega_x = \frac{\Upsilon_x}{\Gamma}, \quad \Omega_\phi =\frac{\Upsilon_\phi}{\Gamma}, \quad \Omega_\psi = \frac{\Upsilon_\psi}{\Gamma}.
\end{aligned}
\end{align}

The perihelion shift and the Lense-Thirring effect may now be calculated by

\begin{align}
\begin{aligned}
\Delta_{\rm P}^\phi &= \Omega_\phi - \Omega_x = \frac{\Upsilon_\phi - \Upsilon_x}{\Gamma},\\
\Delta_{\rm P}^\psi &= \Omega_\psi - \Omega_x = \frac{\Upsilon_\psi - \Upsilon_x}{\Gamma},\\
\Delta_{\rm LT}^\phi &=  \Omega_\phi - \Omega_\theta =  \frac{\Upsilon_\phi -\Upsilon_\theta}{\Gamma},\\
\Delta_{\rm LT}^\psi &=  \Omega_\psi - \Omega_\theta =  \frac{\Upsilon_\psi -\Upsilon_\theta}{\Gamma}.
\end{aligned}
\end{align}

\section{Orbits}

In this section, we show examples of the orbit types discussed in the previous sections. The orbits are plotted in cartesian coordinates $(X, Y, Z, W )$ defined by

\begin{align}
\begin{aligned}
X &= \sqrt{x+2a^2} \, \sin \theta \, \cos \phi,\\
Y &= \sqrt{x+2a^2} \, \sin \theta \, \sin \phi,\\
Z &= \sqrt{x+a^2+b^2} \, \cos \theta \, \cos \psi,\\
W &= \sqrt{x+a^2+b^2} \, \cos \theta \, \sin \psi.
\end{aligned}
\end{align}

\subsection{2D plots}

In the case of $\theta = 0$ or $\theta = \frac{\pi}{2}$, the four-dimensional orbit reduces to a planar orbit in the $X$-$Y$-plane or $Z$-$W$-plane, respectively. We want to consider orbits in the $\theta = \frac{\pi}{2}$ plane, thus we need to choose $\Psi=0$. Furthermore, according to Eq. \eqref{eq:Kthetahalfpi}, the Carter constant $K$ has to be be expressed in terms of 

\begin{align}
K = \Phi^2 - \left(E^2 - \delta \right) b^2.
\end{align}

Since, the geodesic equation of $\phi$ is now only depending on the $x$-coordinate, we may find radial turning points in such a way that $\dot \phi$ changes its sign. The related boundary is called turnaround boundary and was introduced in \cite{Diemer:2013fza} due to the fact the the BMPV spacetime does not possess an ergoregion, since its horizon angular velocity vanishes. In the  $\theta = \frac{\pi}{2}$-plane this equations yields two solutions.

Massive, two-dimensional escape orbit:

\begin{figure}[H]
	\centering
	\captionsetup{justification=justified}
	\includegraphics[width=0.45\textwidth]{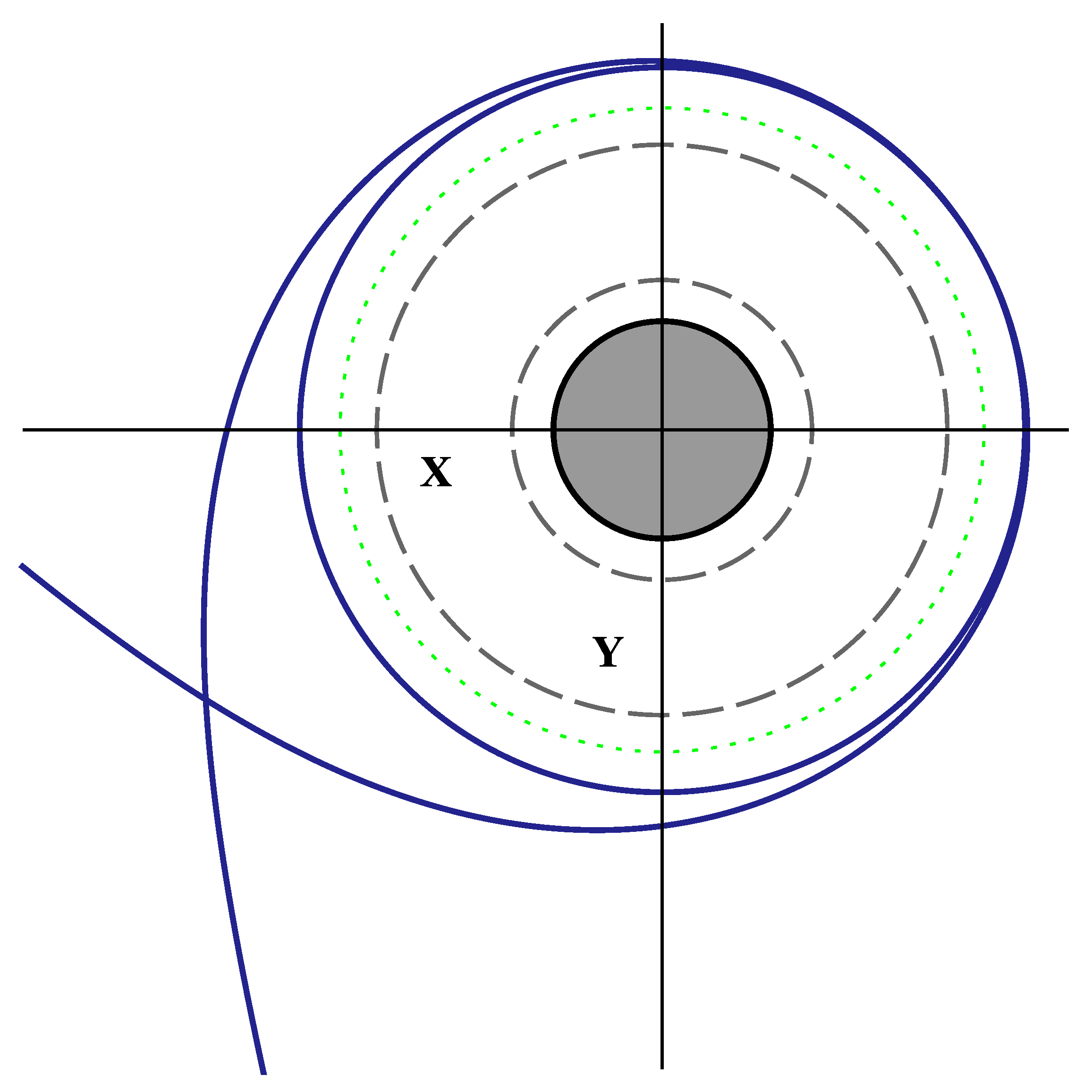}
	\caption{Massive, two-dimensional escape orbit (blue)  in the $X$-$Y$-plane for parameter values: $a = 0.3, b = 0.2, q=0.2, \mu = 1; \Phi = -2, E=1.42274248$. The grey dashed lines represent the horizons and the green dotted line denotes the static limit. The inner static limit is beyond the singularity, which is depicted by the grey disk, and therefore not shown.}
\end{figure}

For the same set of parameter values but a different radial starting point, we will obtain a massive, two-dimensional many-world bound orbit:

\begin{figure}[H]
	\centering
	\captionsetup{justification=justified}
	\includegraphics[width=0.4\textwidth]{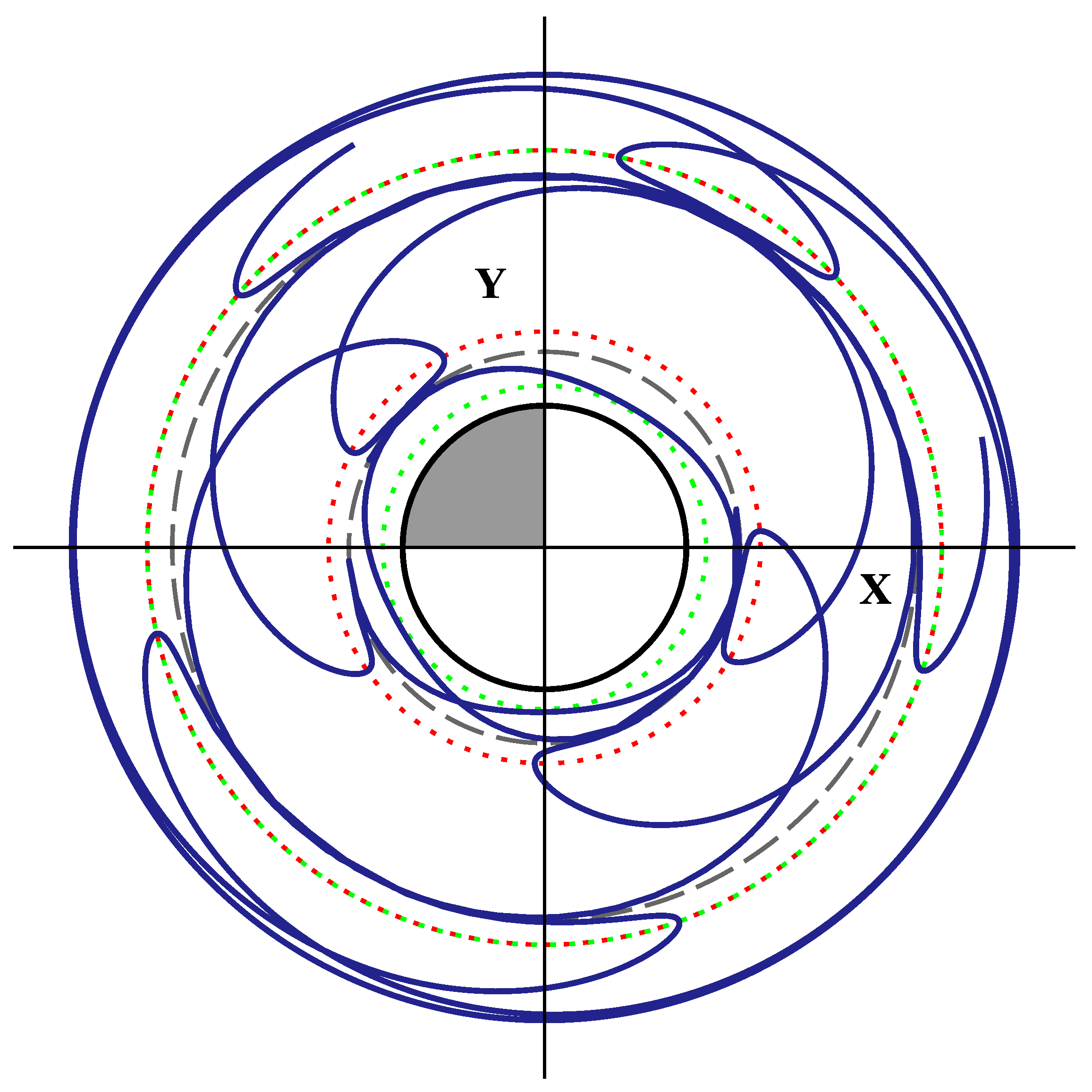}
	\caption{Massive, two-dimensional many-world bound orbit  in the $X$-$Y$-plane for parameter values: $a = 0.3, b = 0.2, q=0.2, \mu = 1; \Phi = -2, E=1.42274248$. Here, both values of the static limit (green dotted lines) and turnaround boundaries (red dotted lines) are shown. Both boundaries seem to merge outside the event horizon, while they are clearly disctinct inside.}
\end{figure}

Furthermore, we present a massive, two-dimensional two-world escape orbit:

\begin{figure}[H]
	\centering
	\captionsetup{justification=justified}
	\includegraphics[width=0.49\textwidth]{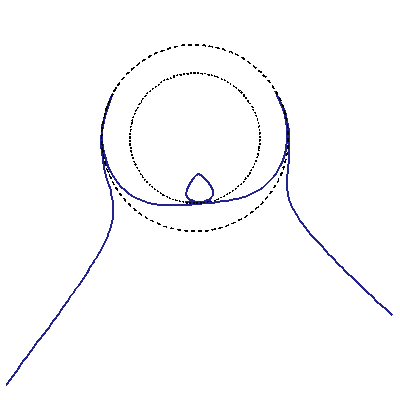}
	\caption{Massive, two-dimensional two-world escape orbit  in the $X$-$Y$-plane for parameter values: $a = 0.4, b = 0.3, q=0.2, \mu = 1; \Phi = -1.5, E=3$.}
\end{figure}

The effect of frame-dragging is visible, as the test particle approaches the black hole with an opposite rotational direction and is forced to co-rotate when crossing the static limit.\\

Finally, we will plot a massless, two-dimensional two-world escape orbit:

\begin{figure}[H]
	\centering
	\captionsetup{justification=justified}
	\includegraphics[width=0.52\textwidth]{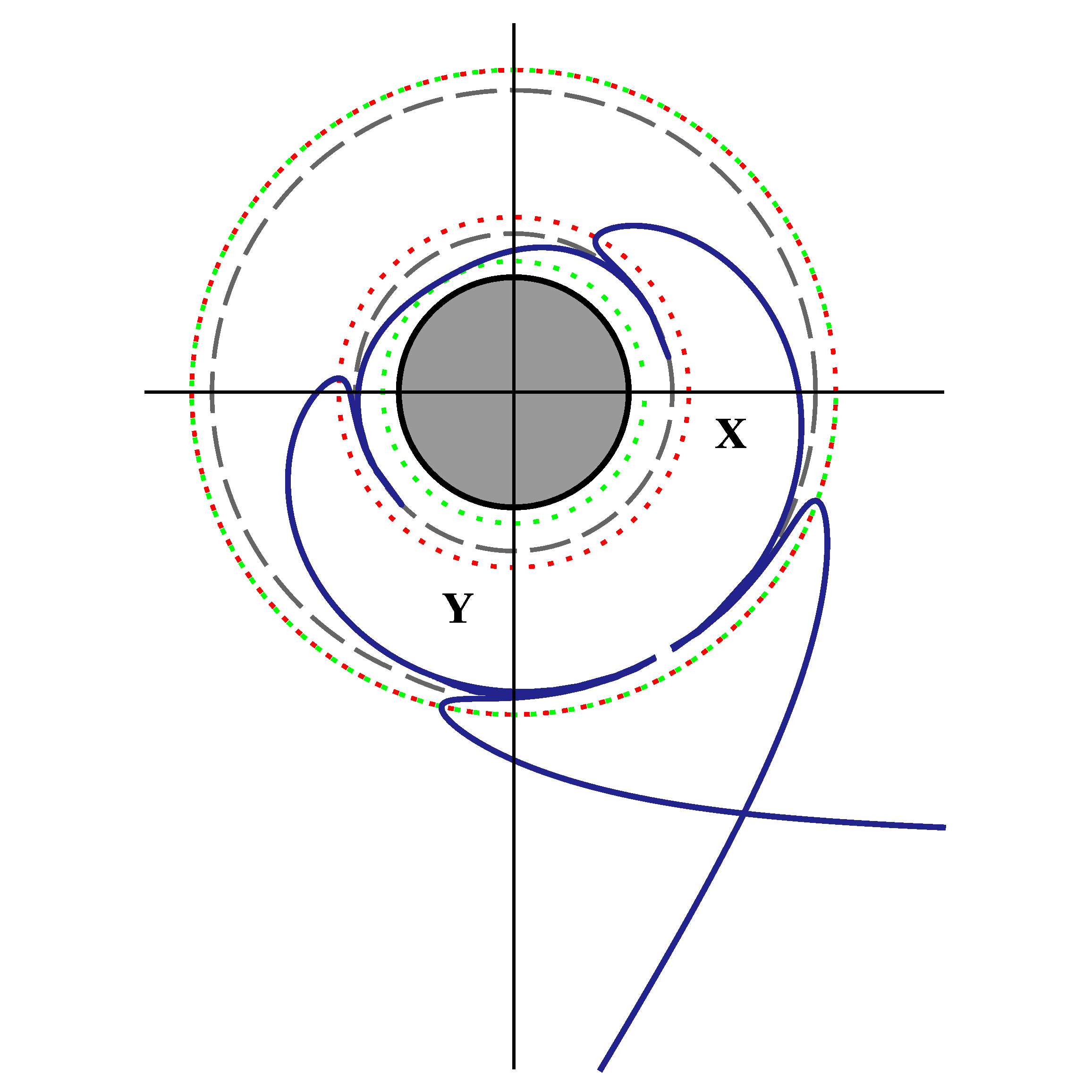}
	\caption{Massless, two-dimensional two-world escape orbit in the $X$-$Y$-plane for parameter values: $a = 0.3, b = 0.2, q=0.2, \mu = 1; \Phi = -2, E=1.42274248$.}
\end{figure}

Instead of restricting the test particle motion to a certain subspace, we may choose hyperslices in the $X$-$Z$-plane by setting $\psi = \phi = const. = 0$. Of course, this does not represent a physical two-dimensional geodesic as we have shown in the former plots, but a two-dimensional slice of the actual orbital motion. In this manner, a massive escape orbit is shown in Fig. \ref{fig:EOXZ}:

\begin{figure}[H]
	\centering
	\captionsetup{justification=justified}
	\includegraphics[width=0.49\textwidth]{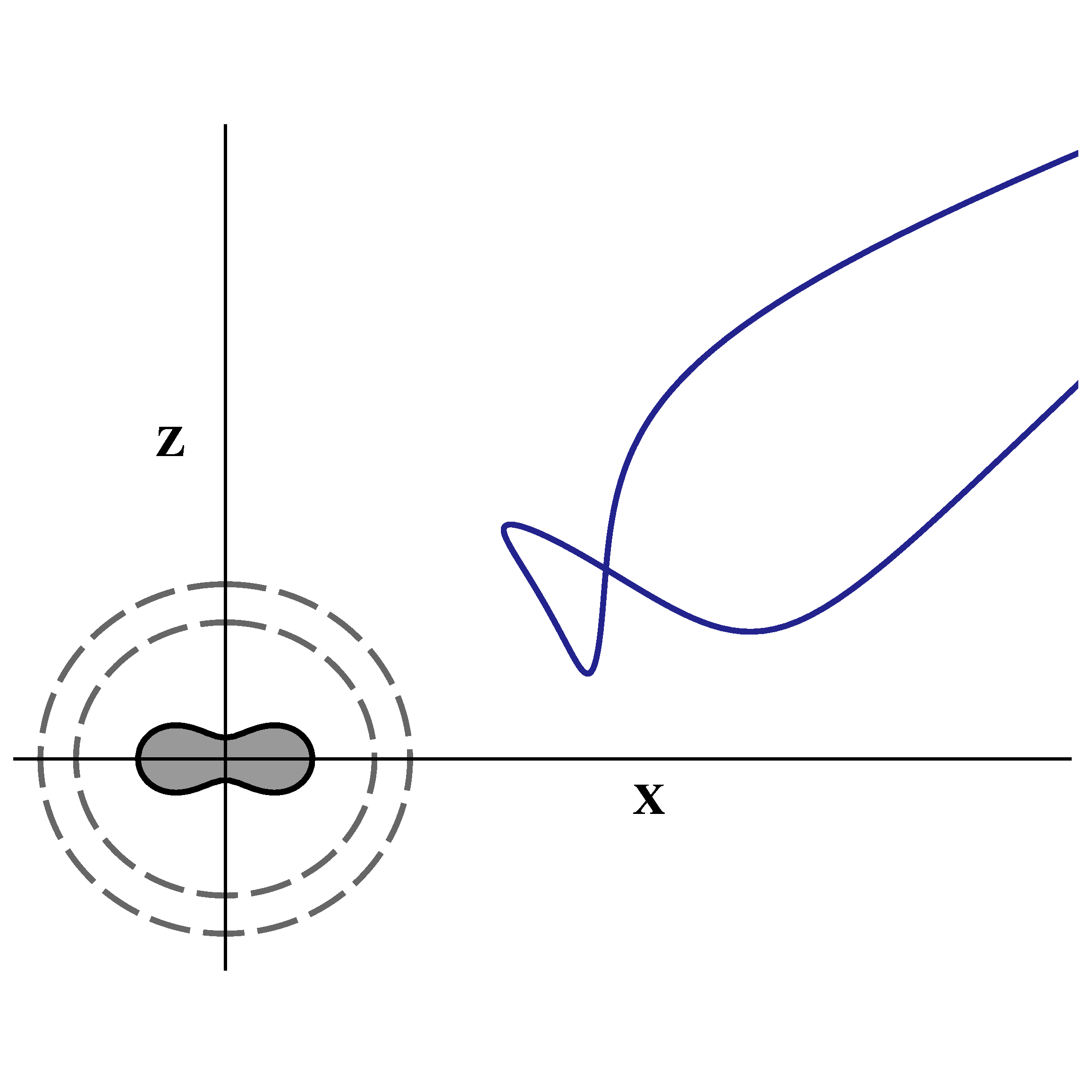}
	\caption{Massive escape orbit in the $X$-$Z$-plane for parameter values: $a = 0.3, b = 0.1, q=0.4, \mu = 1; \Phi = -1, \Psi = -0.2, K=1.8, E=1.09$.}
	\label{fig:EOXZ}
\end{figure}

Since our chosen spacetime hyperslices do neither consider a $\phi$- nor a $\psi$-motion anymore, the divergences at the horizons are omitted. This can be seen in Fig. \ref{fig:MBOXZ}, which represents a massive many-world bound orbit:

\begin{figure}[H]
	\centering
	\captionsetup{justification=justified}
	\includegraphics[width=0.49\textwidth]{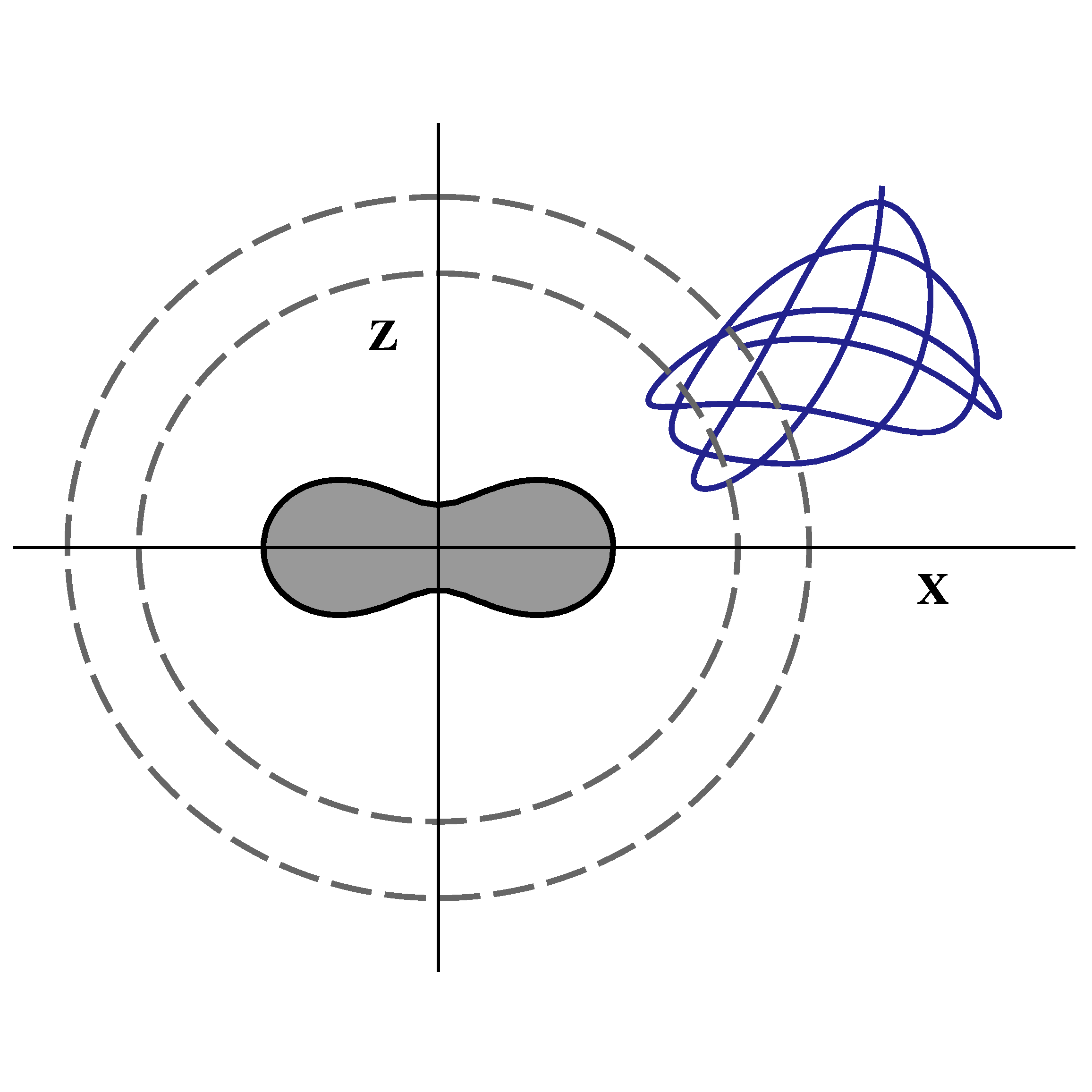}
	\caption{Massive many-world bound orbit in the $X$-$Z$-plane for parameter values: $a = 0.3, b = 0.1, q=0.4, \mu = 1; \Phi = 1, \Psi = -0.2, K=1.8, E=0.9$.}
\label{fig:MBOXZ}
\end{figure}

Fig. \ref{fig:MBOXZlight} depicts a massless many-world bound orbit: 

\begin{figure}[H]
	\centering
	\captionsetup{justification=justified}
	\includegraphics[width=0.49\textwidth]{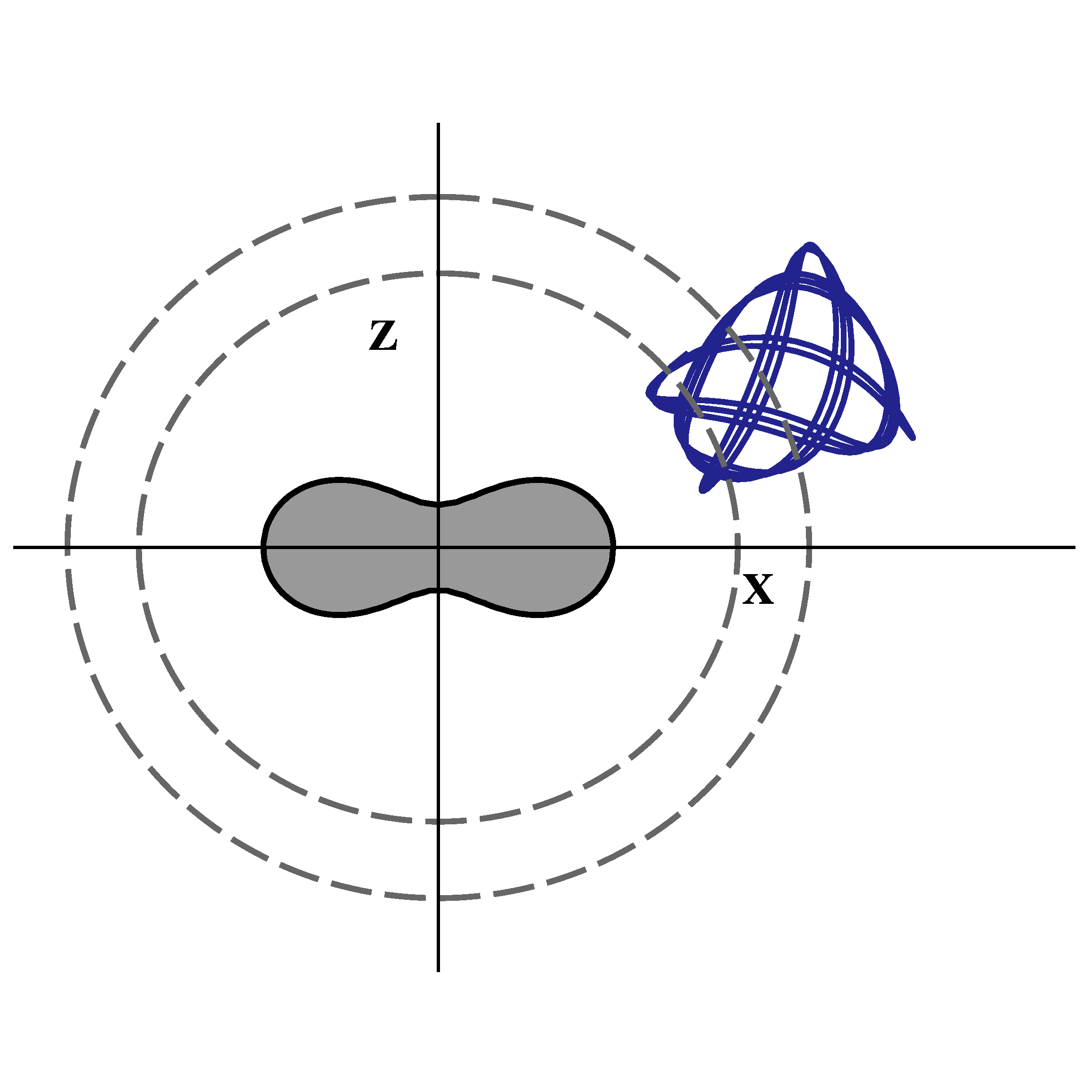}
	\caption{Massless many-world bound orbit in the $X$-$Z$-plane for parameter values: $a = 0.3, b = 0.1, q=0.4, \mu = 1; \Phi = -1, \Psi = -0.2, K=1.8,  E=0.5$.}
\label{fig:MBOXZlight}
\end{figure}

A massive two-world escape orbit is shown in Fig. \ref{fig:TEOXZ}:

\begin{figure}[H]
	\centering
	\captionsetup{justification=justified}
	\includegraphics[width=0.49\textwidth]{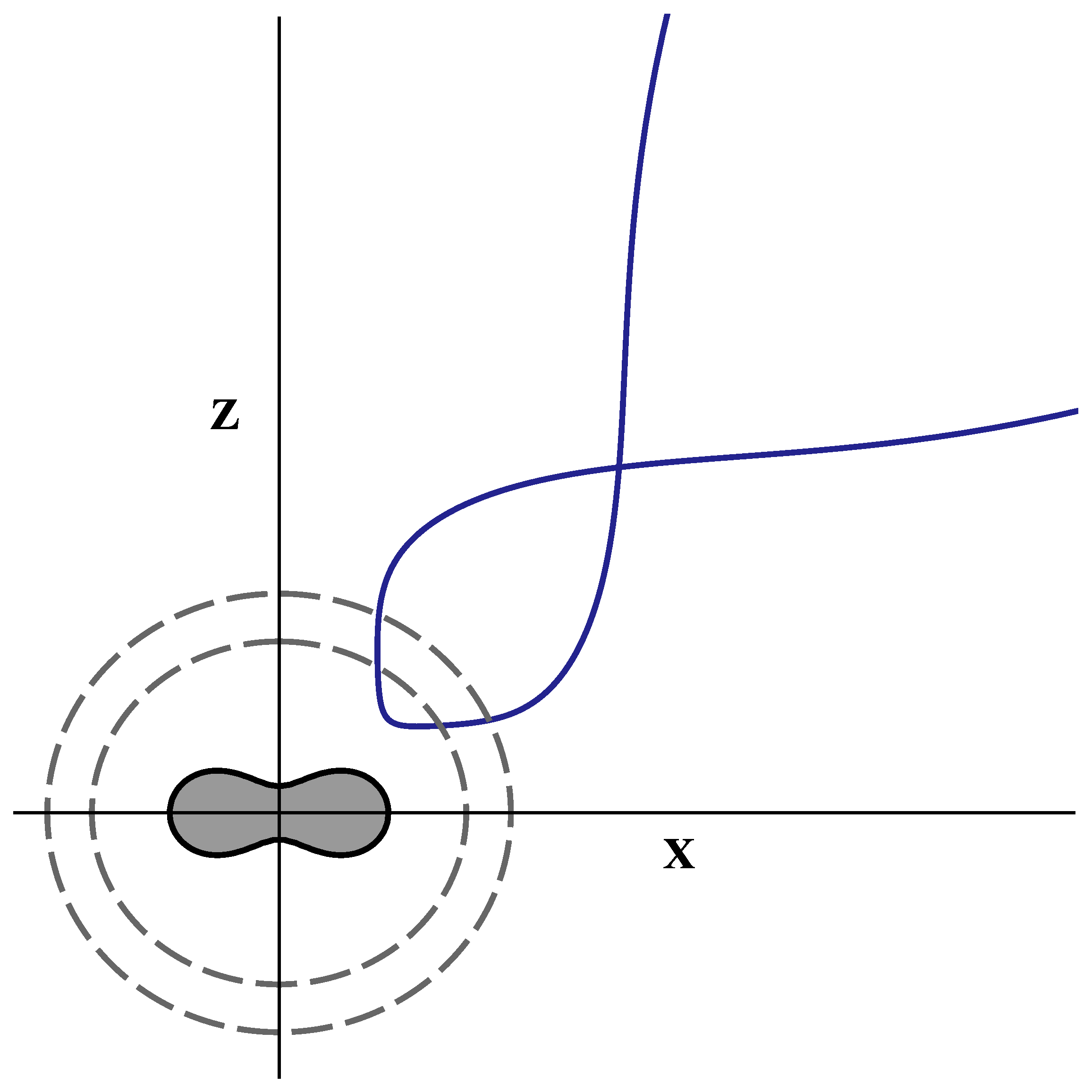}
	\caption{Massive two-world escape orbit in the $X$-$Z$-plane for parameter values: $a = 0.3, b = 0.1, q=0.4, \mu = 1; \Phi = 0.5, \Psi = 0.5,K=1.8, E=1.2$.}
\label{fig:TEOXZ}
\end{figure}

\subsection{3D plots}

In order to obtain three-dimensional representations of the test particle motion, we simply omit one cartesian coordinate (e.g. the $W$-coordinate), which produces a projection of the orbital motion.

\begin{figure}[H]
	\centering
	\captionsetup{justification=justified}
		\includegraphics[width=0.49\textwidth]{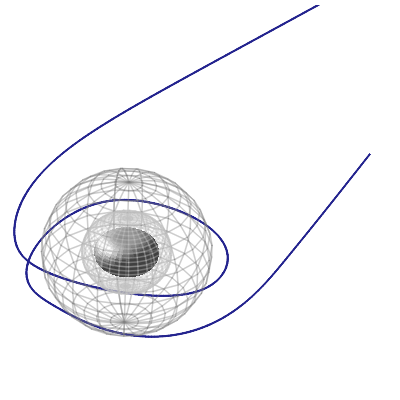}
\caption{Massive, three-dimensional projection of an escape orbit in the $X$-$Y$-$Z$-space for parameter values: $a = 0.3, b = 0.2, q=0.2, \mu = 1; \Phi = -2, \Psi = -0.2, K=5,  E=1.5549783$.}
\end{figure}

\begin{figure}[H]
	\centering
	\captionsetup{justification=justified}
	\includegraphics[width=0.42\textwidth]{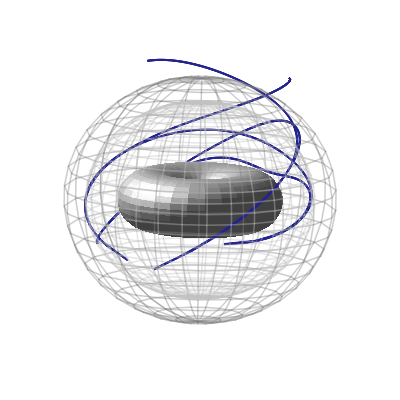}
\caption{Massive, three-dimensional projection of a many-world bound orbit in the $X$-$Y$-$Z$-space for parameter values: $a = 0.4, b = 0, q=-0.4, \mu = 1; \Phi = 0.6, \Psi = -0.2, K=1.8,  E=0.5$.}
	\end{figure}
	
	\begin{figure}[H]
		\centering
		\captionsetup{justification=justified}
		\includegraphics[width=0.5\textwidth]{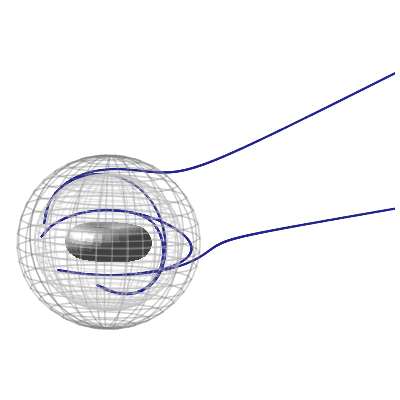}
\caption{Massive, three-dimensional projection of a two-world escape orbit in the $X$-$Y$-$Z$-space for parameter values: $a = 0.3, b = 0.1, q=0.4, \mu = 1; \Phi = -1, \Psi = -0.2, K=1.8,  E=3$.}
	\end{figure}

\section{Outlook}

In this paper we have discussed the motion of test particles in the five-dimensional, charged, rotating Einstein-Maxwell-Chern-Simons spacetime. We derived the geodesic equations of motion, studied their general properties analyzed the structure of
the resulting orbits. Therefore, we investigated the effective potentials of the radial and polar motion and employed them in order to classify the possible types of orbits in this spacetime.\\ 
We integrated the equations of motions analytically in terms of the Weierstrass elliptic $\wp$-, $\zeta$- and $\sigma$-functions. We presented the analytical expressions for light deflection, the periastron shift and the Lense-Thirring effect. Finally we used the analytical solutions in order to visualize the orbital motion in two- and three-dimensional plots.\\
Throughout, we emphasized the influence of the black hole's charge and elaborated the similarities and differences to related spacetimes e.g. the five-dimensional Myers-Perry spacetime or the four-dimensional Kerr-Newman spacetime. Due to the fact that the spacetime action contains a Chern-Simons term besides the usual Maxwell term, the sign of the black hole's charge has an influence on the spacetime structure e.g. the location of the horizons. Similar to the five-dimensional Myers-Perry spacetime, we did not find any bound orbits outside the event horizon but exclusively hidden behind the Cauchy horizon. This type of orbit is also known in the Reissner-Nordström spacetime \cite{Grunau:2010gd}.\\
As a future step, it would be interesting to add an electric charge to the test particle, although it is not known if the equations of motion are still separable, yet. There may be bound orbits outside the event horizon for some appropriate value of the black hole's and test particle's electric charge. Furthermore, a supplemental cosmological constant would have a big influence on the spacetime structure and the possible types of orbits. Accordingly, this spacetime could be compared to the five-dimensional Myers-Perry spacetime with cosmological constant (under preparation). The addition of a cosmological constant in five-dimensional spacetimes is also interesting concerning the AdS/CFT correspondence.\\
The analysis of other charged rotating spacetimes in higher dimensions, such as charged rotating black holes in Einstein-Maxwell-dilaton theory \cite{Kunz:2006jd}, might be interesting, as well.

\section*{Acknowledgement}

We gratefully acknowledge support by the Deutsche Forschungsgemeinschaft (DFG), in particular, within the framework of the DFG Research Training group 1620 {\it Models of gravity} and the Working Internships in Science and Engineering (WISE) by Deutscher Akademischer Austauschdienst (DAAD).

\newpage
\hspace{1cm}
\newpage

\bibliographystyle{unsrt}

\end{document}